\documentclass[11pt]{article}

\usepackage{times}
\usepackage{graphicx}
\usepackage{color}
\usepackage{amsmath, amsfonts, amssymb}
\usepackage{ccaption}
\usepackage{verbatim}
\usepackage{mathtools}
\usepackage{versions}
\usepackage{xr}
\usepackage{booktabs}
\usepackage[labelfont=bf]{caption}
\usepackage [autostyle, english = american]{csquotes}
\MakeOuterQuote{"}
\usepackage{setspace}

\title{Stochastic Failure Dynamics in a Model
Peptidoglycan Mesh under Hydrolysis and Mechanical Loading}

\author
{Issan Patri$^{1,*}$, Garima Rani$^{2,*}$ \\
\\
\normalsize{$^{1}$Theoretical Statistics and Mathematics Unit, Indian Statistical Institute, Delhi 110016, India}\\
\normalsize{$^{2}$Department of Biochemical Engineering and Biotechnology, Indian Institute of Technology,}\\
\normalsize{Hauz Khas, Delhi 110016, India}\\
\normalsize{$^{*}$ Corresponding authors: issanp@isid.ac.in, garimarani@iitd.ac.in}\\
}

\date{}

\begin{document}

\maketitle

\begin{abstract}
We develop a minimal stochastic-mechanical model for failure dynamics of the Gram-negative peptidoglycan (PG) cell wall under blocked growth and continued hydrolysis. The cell wall is modelled as a collection of load-bearing cross-sections, with random number of cross-links in each such cross-section, around a fixed degree of cross-linking. To understand failure dynamics of the cell wall under antibacterial agents which prevent insertion of fresh PG material, peptide bonds are progressively excised by hydrolases, while insertion/formation of new bonds is blocked. We compare and contrast rupture dynamics in no-load and load-activated regimes, using extreme value theory and stochastic simulations, deriving limiting laws for system failure in both the long-cell and large-crosslinking limits. For a cross-section with many bonds, the mean failure time grows only logarithmically with bond number, giving diminishing returns from additional cross-linking. In the no-load case, the centered failure time converges to a Gumbel law, though crucially, load changes the centered distribution and accelerates the final stages of collapse. In long cells, failure follows a weakest-link Weibull-type scaling controlled by the least cross-linked regions, with mechanical load reducing the failure-time scale without changing the weakest-link exponent. The fraction of bonds removed before failure tends to zero, showing that collapse can remain highly localized even while the global cross-linking degree stays high. Slowing down growth and other cellular functions is a typical antibiotic escape mechanism for bacteria and in this context, we show that, hydrolase downregulation uniformly rescales the failure-time distribution and delays collapse, but does not remove the load-induced preference for rupture in weakened regions.

\end{abstract}

\section{\label{sec:level1}INTRODUCTION}

Bacteria are one of the most successful organisms on earth, having adapted to thrive in a diverse range of habitats, from human gut to hot springs to nuclear waste. A key component of their evolutionary resilience is the bacterial cell wall, a remarkable load-bearing structure which resists turgor pressure and is a key determinant of cell shape \cite{VollmerHoltje2004,Huang2008,Garde2021,Turner2018,Deghelt2025}. In Gram-negative rod-shaped bacteria such as \textit{Escherichia coli}, the peptidoglycan (PG) layer is a principal load-bearing component of the cell envelope, comprising a thin mesh of glycan strands cross-linked by short peptide bridges \cite{Huang2008,Garde2021,Vollmer2008}. The PG mesh must satisfy diverse competing mechanical demands- it must be strong enough to resist the large internal turgor pressure, sufficiently elastic to allow growth and remodeling, and tough enough to avoid catastrophic rupture under defects or damage \cite{Huang2008,Deghelt2025,Gumbart2014,Rojas2018Nature,RaniPatri2019}.

A distinctive feature of the PG layer is its heterogeneous ultrastructure, comprising glycan strands that have a broad length distribution, cross-links which are present intermittently at a fraction of possible peptide stems, with their local placement varying across the sacculus \cite{Vollmer2008,Gan2008,RaniPatri2020,Alvarez2024}. Strikingly, the PG sacculus is also a highly dynamical structure. During growth, old bonds are continuously cleaved by hydrolases while new glycan strands and peptide cross-links are inserted, so the wall is maintained through a balance of degradation and repair \cite{UeharaBernhardt2011,Egan2017,Egan2020,Zhang2023}. This continual turnover is mechanically delicate, as the PG sacculus is under turgor pressure of order $\sim 1$ atm \cite{Huang2008,Rojas2018Nature,Cayley2000}.

The mechanical organization and remodelling of the PG layer has motivated several experimental studies and theoretical models. Pioneering work by Koch emphasized the problem of how bacteria can grow and divide in spite of internal hydrostatic pressure, and introduced the idea that wall expansion requires a careful coordination between bond cleavage and insertion of new material \cite{Koch1983,Koch1985,Holtje1998,Koch2000}. Later mechanical and biophysical models have approached the bacterial cell wall from a variety of complementary viewpoints. For instance, Huang et al. treated the Gram-negative cell wall as a pressurized elastic spring network and showed how local patterns of peptidoglycan defects can generate bulging, cracking, and changes in cell shape \cite{Huang2008}. Amir, Nelson and co-workers developed a dislocation-mediated picture of rod-shaped cell-wall growth, connecting the motion of wall-synthesis complexes, turgor-induced stress, and the cylindrical geometry of the sacculus \cite{AmirNelson2012}. Other studies have shown that growing bacterial cell walls can deform plastically under sustained bending forces, indicating that mechanical stress can regulate wall growth and shape recovery \cite{AmirJun2014}. Other continuum and coarse-grained approaches have connected wall elasticity, biochemical growth, cross-link density, multilayer peptidoglycan architecture, and global cell-shape dynamics \cite{Gumbart2014,RaniPatri2019,RaniPatri2020, BanerjeeSchererDinner2016}. More recently, the cell envelope has been emphasized as an active mechanical structure, linking envelope stiffness, curvature, nonlinear wall mechanics, turgor regulation, and load sharing with wall components to robust bacterial shape control, as well as connecting cell-wall synthesis, hydrolase organization, and repair of wall defects to bacterial morphogenesis and mechanical homeostasis \cite{Deghelt2025, Rojas2018Nature,Alvarez2024,AlMoslehRojas2022,BanzhafVanTeeffelen2020,VigourouxVanTeeffelen2020,OldewurtelVanTeeffelen2023}.

Previous work has therefore studied several aspects of PG mechanics, including the relationship between PG mesh geometry and material properties, the coordination of wall growth with local remodeling, and the role of mechanical feedback in maintaining bacterial shape \cite{Huang2008,Gumbart2014,RaniPatri2019,RaniPatri2020,Egan2020,Zhang2023}. However, the failure dynamics of the PG mesh remain less well understood, particularly when one or more steps in the normal remodeling cycle are naturally or pharmacologically disrupted.

In general in engineering, failure-mode analysis is a standard way of understanding why structures work and how they should be designed. Classical fracture mechanics, beginning with Griffith's theory of crack propagation, asks when small defects become unstable cracks \cite{Griffith1921}. Fatigue analysis studies failure under repeated loading, buckling theory studies sudden collapse under compression, creep rupture studies slow failure under sustained load, and fiber-bundle models describe progressive collapse under redistributed forces \cite{Pradhan2010}. Such approaches have been central to the design and analysis of aircraft, pressure vessels, bridges, composite materials, and thin shells. By contrast, biological materials are often studied primarily through their normal function: how they grow, remodel, bear load, or generate shape. Yet biological structures are also constrained by failure. The bacterial cell wall is a particularly important example. It must remain mechanically intact while being continuously remodeled by enzymes that cut and rebuild the peptidoglycan mesh \cite{UeharaBernhardt2011,Egan2020,Zhang2023}. Therefore, for a complete understanding of the design principles of the cell wall, it is imperative to understand how it fails, what fail-safes it possesses, and how cells avoid catastrophic collapse when the remodelling process is disrupted. This motivates the present work- we study the failure modes of the PG mesh in a regime where new wall insertion is blocked while hydrolytic cleavage continues.

This question is directly relevant to antibiotic action. Given the central role of the cell wall in promoting cell survival, it is not surprising that the PG layer is a prime target of anti-bacterial agents, starting from the serendipitous discovery of the first antibiotic, penicillin. In particular, multiple antibacterial agents target cell-wall synthesis by blocking the production, insertion, or cross-linking of new PG material \cite{Garde2021,Egan2020}. Antibiotics such as beta-lactams, glycopeptides such as vancomycin, fosfomycin, bacitracin, and D-cycloserine interfere with different stages of peptidoglycan synthesis, from precursor formation to transport, glycan polymerization, and peptide cross-linking \cite{Garde2021}. These drugs can motivate the type of regime considered here, in which addition of new wall material is strongly impaired while hydrolytic cleavage of existing bonds can continue \cite{OldewurtelVanTeeffelen2023,ChoUeharaBernhardt2014}. In such a situation, the mechanical fate of the cell is not determined only by synthesis inhibition. It is also determined by the continued action of hydrolases, which cleave existing peptide or glycosidic bonds during normal wall remodeling \cite{UeharaBernhardt2011,Egan2020,Zhang2023}. If insertion is stopped while hydrolysis continues, the wall is progressively degraded. Understanding the failure dynamics of the PG mesh in this regime is therefore important both for understanding antibiotic-induced lysis and for identifying possible design principles for new antibacterial agents \cite{ChoUeharaBernhardt2014}.

To address this, we develop a minimal stochastic-mechanical model of peptidoglycan failure in which insertion of new wall material is blocked while hydrolytic cleavage of peptide cross-links continues. The PG mesh is coarse-grained into a sequence of load-bearing cross-sections containing random numbers of peptide bonds, with system failure defined by the first cross-section to become fully depleted. We first study failure in the no-load regime and then incorporate mechanical loading due to turgor pressure through force sharing among the surviving bonds of each cross-section. The model reveals two distinct sources of mechanical vulnerability. At the local level, increasing the number of peptide bonds provides only logarithmically increasing protection: in the many-bond limit the centered no-load failure time converges to a Gumbel law, whereas mechanical loading preserves the leading logarithmic scale but produces a distinct load-dependent fluctuation law which accelerates collapse of depleted regions. At the system level, increasing the number of cross-sections creates a weakest-link effect: in the long-cell limit failure time follows a Weibull-type scaling controlled by the less cross-linked regions, while load reduces the corresponding failure-time scale without changing the weakest-link exponent \cite{Pradhan2010, Weibull1951}. Failure can therefore occur after only a vanishing fraction of the total bonds has been removed, so that a locally failed cross-section may coexist with a globally highly cross-linked mesh. Cross-linking protects both by increasing frequency of load bearing peptides locally and by suppressing rare weak regions. On the other hand, hydrolase downregulation acts differently, by uniformly rescaling the failure-time distribution in time without altering the load-dependent spatial pathway of rupture. Our model and results thus highlight how effects such as local heterogeneity in the PG mesh, evolution of system size, mechanical load, and activity drive PG failure under blocked wall synthesis and allow for separating these effects.

\section{\label{sec:level2}MODEL}

\begin{figure}[t]
\centering
    \includegraphics[width= 5 in]{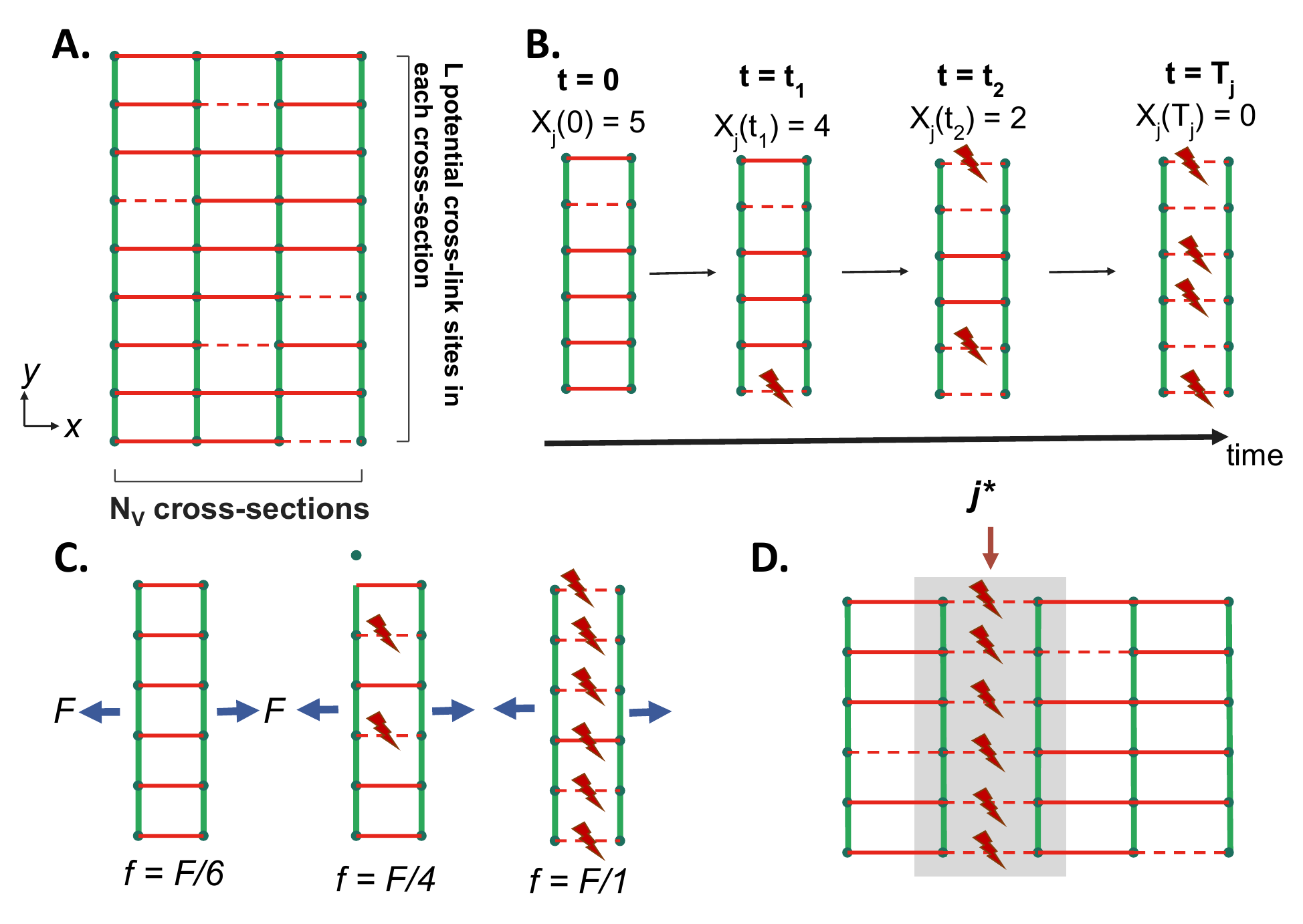}
    \caption{\textbf{Stochastic-mechanical model of peptidoglycan mesh failure.}
A. PG mesh represented in the $x$--$y$ plane with glycan strands connected by peptide cross-links at $L$ potential cross-linking sites per cross-section. Cross-links occur with probability $p$.
B. Stochastic hydrolysis of a single cross-section $j$, so the number of intact peptide cross-links $X_j(t)$ decreases from its initial value to zero. Local failure occurs at $T_j$ when $X_j(T_j)=0$.
C. Force redistribution under a fixed load $F$, as peptide cross-links are cleaved, the load is equally shared among fewer surviving bonds.
D. Each cross-section has a stochastic failure time, and the PG mesh fails when the first cross-section becomes fully depleted. The failed cross-section $j^*$ is highlighted.}
    \label{fig:schematic}
\end{figure}

We model the PG layer as a square lattice with glycan strands aligned vertically and the peptide cross-links aligned horizontally. Thus, the PG layer is visualised as a series of a fixed number of circumferential cross-sections (say $N_V$), each such cross-section comprising of $L$ sites for peptide bonds. However, not all sites have a peptide bond and the number of bonds is determined by the degree of crosslinking. Let $p\in[0,1]$ be the crosslinking degree, then at time $t=0$, each cross-section $i$ will have $X_i(0)$ bonds, where $X_i(t)$ denotes the number of bonds in the $i$-th cross-section and for all $i$, $X_i(0)$ is taken from independent draws from a binomial distribution Binomial$(L,p)$. Thus, each cross-section will have on an average $pL$ peptide bonds at $t=0$ which are arranged in parallel (Fig. \ref{fig:schematic}). 

To model failure modes under the action of peptide hydrolases, for each cross-section $i$, we define the local failure time $T_i$ as the time $t\geq 0$ when $X_i(t)=0$, where $X_i(t)$ denotes the number of intact peptide bonds in cross-section $i$ at time $t$. Thus, $T_i$ is the first time at which all peptide bonds in the $i$-th cross-section have been removed. The system fails when the first cross-section loses all of its peptide bonds, so that $T_{\mathrm{sys}}=\min_{1\leq i\leq N_V}T_i$. Thus, the PG mesh need not be globally degraded for mechanical failure, rather a single locally disconnected cross-section will signify system failure (see Fig.\ref{fig:schematic}).

Here, our focus is on PG degradation in which new wall synthesis is arrested but hydrolytic action on peptides continues. Thus, peptide bonds are cleaved from existing cross-sections, but no new peptide bonds are inserted and no new cross-sections are added. This assumption isolates the mechanical consequences of continued PG turnover when the normal balance between synthesis and degradation is disrupted. As noted, such a regime is relevant to situations in which wall synthesis is inhibited, for instance by antibiotics that block the incorporation or cross-linking of new cell-wall material. In normal growth, hydrolysis is coupled to insertion and remodeling \cite{Egan2020, Holtje1998}, while here, we study PG layer dynamics when the degradative part of this process continues in the absence of compensating repair. In terms of the state variables $X_i(t)$, we have that they are non-increasing in time, so, $X_i(t_2)\leq X_i(t_1), 0\leq t_1\leq t_2$. 

We consider two regimes for bond removal- the first is a no-load regime, in which each intact peptide bond is removed independently at a constant hydrolytic rate per bond, corresponding to the assumption that locally, all bonds have access to a similar level of hydrolases (see Appendix A for more details). This represents the case where the PG layer is assumed to be under no load, and bond cleavage is governed only by enzymatic activity. The second is the load-activated regime, in which the PG layer is taken to be under load due to turgor pressure as present in bacterial cells, in which case the rupture rate depends also on the force carried by the surviving bonds (see Appendix B for more details). The peptide bonds in each cross-section share the load equally. As the number of intact peptide bonds decreases in a cross-section, the per bond load increases and rupture rate therefore increases as the cross-section becomes progressively weakened. In both regimes, we assume that hydrolases are uniformly available across the PG mesh, so that spatial differences in failure arise from the local bond configuration rather than from spatial variation in enzyme abundance. The distinction is that, in the no-load case, this uniform hydrolytic activity gives each intact bond the same cutting rate, whereas in the load-activated case the same hydrolytic activity is modulated by the force borne by the surviving bonds, resulting in local heterogeneities (see Appendix A and B).

These two regimes separate two possible routes to failure. The no load case serves as a baseline in which failure is governed only by the geometry of the PG mesh and the stochastic removal of bonds. On the other hand, in the load-activated case, hydrolysis and mechanics are coupled: each bond removal increases the load on the remaining bonds, which can accelerate subsequent rupture and lead to rapid local collapse. Comparing them allows us to distinguish intrinsic design features of the network from additional acceleration caused by force redistribution. 

Necessary geometric, mechanical, and kinetic parameter values are summarized in Table~\ref{tab:parameters}.
    
\begin{table}[ht]
\centering

\begin{tabular}{@{}l l l p{5.2cm}@{}}
\toprule
\textbf{Parameter} & \textbf{Symbol} & \textbf{Value} & \textbf{Source} \\
\midrule

Crosslinking degree         & $p$                   & $0.30$               & \cite{Vollmer2008} \\[2pt]
Cell radius                  & $R$                   & $0.5~\mu\text{m}$    &  \cite{VollmerHoltje2004, Huang2008} \\[2pt]
Cell length                   & $\ell_{\text{cell}}$  & $2.0~\mu\text{m}$    & \cite{VollmerHoltje2004, Huang2008} \\[2pt]
Disaccharide spacing          & $d_{\text{disacc}}$   & $1.03~\text{nm}$     & \cite{Turner2018, VollmerBlanotDePedro2008} \\[2pt]
Axial hoop spacing            & $d_{\text{ax}}$       & $4.5~\text{nm}$      & \cite{Turner2018, Gan2008} \\[2pt]
Turgor pressure                & $P$                   & $1~\text{atm}$       & \cite{Cayley2000} \\[2pt]
PG turnover per generation    & --                    & $45\%$               & \cite{park2001} \\[2pt]
Generation time                & $T_{\text{gen}}$     & $30~\text{min}$      & \cite{Onogi2002} \\
Activation distance            & $x^{\ddagger}$        & $0.2~\text{nm}$       & representative value \cite{Bell1978,EvansRitchie1997}\\[2pt]\\[2pt]
Thermal energy                  & $k_BT$                & $4.1~\text{pN}\cdot\text{nm}$ & \cite{phillips2012physical} \\[2pt]
Sites per cross-section & $L = \dfrac{2\pi R}{d_{\text{disacc}}}$
  & $\approx 3050$ & - \\[8pt]
Number of cross-sections & $N_V = \dfrac{\ell_{\text{cell}}}{d_{\text{ax}}}$
  & $\approx 444$ & - \\[8pt]
Reference hydrolytic rate & $r_0 = \dfrac{-\ln(1-0.45)}{T_{\text{gen}}}$
  & $\approx 3.32\times10^{-4}~\text{s}^{-1}$ & - \\[8pt]
Physiological load parameter & $\alpha_{\text{phys}} = \dfrac{P\pi R^2 x^{\ddagger}}{k_BT}$
  & $\approx 3882$ & - \\
\bottomrule
\end{tabular}

\caption{Parameters used in the stochastic-mechanical model.}
\label{tab:parameters}

\end{table}

\section{Results}

\subsection*{Failure under hydrolytic action without load}

We first study the baseline case in which peptide bonds are removed by hydrolysis but rupture is not mechanically activated, with the PG mesh taken to be under no load (visualised as large fragment of PG mesh under the action of hydrolytic enzymes). The aim is to isolate the effect of the PG mesh architecture and stochastic bond cutting, to compare and contrast with force-dependent acceleration of hydrolytic cleavage of PG mesh under load.

As noted, hydrolases are assumed to be uniformly available across the PG mesh at a fixed local concentration. Consequently, in physical
time, each intact peptide bond is cut independently at the same rate
$r_0$. Rescaling time by $r_0$ gives dimensionless time $t$, in which
the lifetime of each bond is an independent exponential random
variable with unit rate. Here, $\operatorname{Exp}(\lambda)$ denotes
an exponential distribution with rate $\lambda$, whose mean is
$1/\lambda$. 

Consider a cross-section initially containing $k$ peptide bonds.
Let $\xi_1,\ldots,\xi_k$ denote the times at which these bonds are
removed. As noted, failure in our model occurs when the last
surviving bond in the cross-section is removed, thus its failure time
is the largest of these $k$ independent waiting times, giving
\begin{equation}
T_k^{(0)}
=
\max\{\xi_1,\dots,\xi_k\},
\qquad
\xi_j\sim \operatorname{Exp}(1).
\end{equation}

The same process can also be described by following successive
bond-removal events. When $m$ bonds remain, each has unit cutting
rate, so the total rate of the next cleavage event is $m$.
The waiting time until that event therefore follows
$\operatorname{Exp}(m)$ and so we get
\begin{equation}
T_k^{(0)}
=
\sum_{m=1}^{k}\tau_m,
\qquad
\tau_m\sim\operatorname{Exp}(m),
\end{equation}
where the $\tau_m$ are the independent waiting times between
successive cleavage events. The equality here holds since the failure time is the sum of all waiting times until the last bond is removed. Taking expectations we get
\begin{equation}
\mathbb{E}\!\left[T_k^{(0)}\right]
=
\sum_{m=1}^{k}\frac{1}{m}
=
H_k
\approx \log k+\gamma.
\end{equation}
where $H_k=\sum_{m=1}^{k}\frac{1}{m}$ and $\gamma$ is Euler's constant (thus, in physical time, we have $\mathbb{E}\!\left[T_{k,phys}^{(0)}\right]\approx \frac{\log k+\gamma}{r_0}$). This logarithmic dependence, even at the level of a single cross-section, is an indication that cross-linking provides only a slowly increasing safety margin against hydrolytic failure in general. So, while adding more peptide bonds to a cross-section indeed increases its expected lifetime, but importantly, the gain is sublinear: increasing the initial bond number from $k$ to $ck$ changes the expected failure time by only $\sim \log c$.

To model the variability in the actual number of cross-linking peptide bonds across the PG mesh, we take the initial number of bonds in each cross-section from a binomial distribution. So, we take $K_i=X_i(0)\sim \operatorname{Binomial}(L,p)$ for $i=1,2,\dots, N_V$, which gives the typical initial number of bonds in a cross-section to be $pL$, $p$ thus defining the degree of cross-linking in the PG mesh. 

\begin{figure}[h]
\centering
    \includegraphics[width= 5 in]{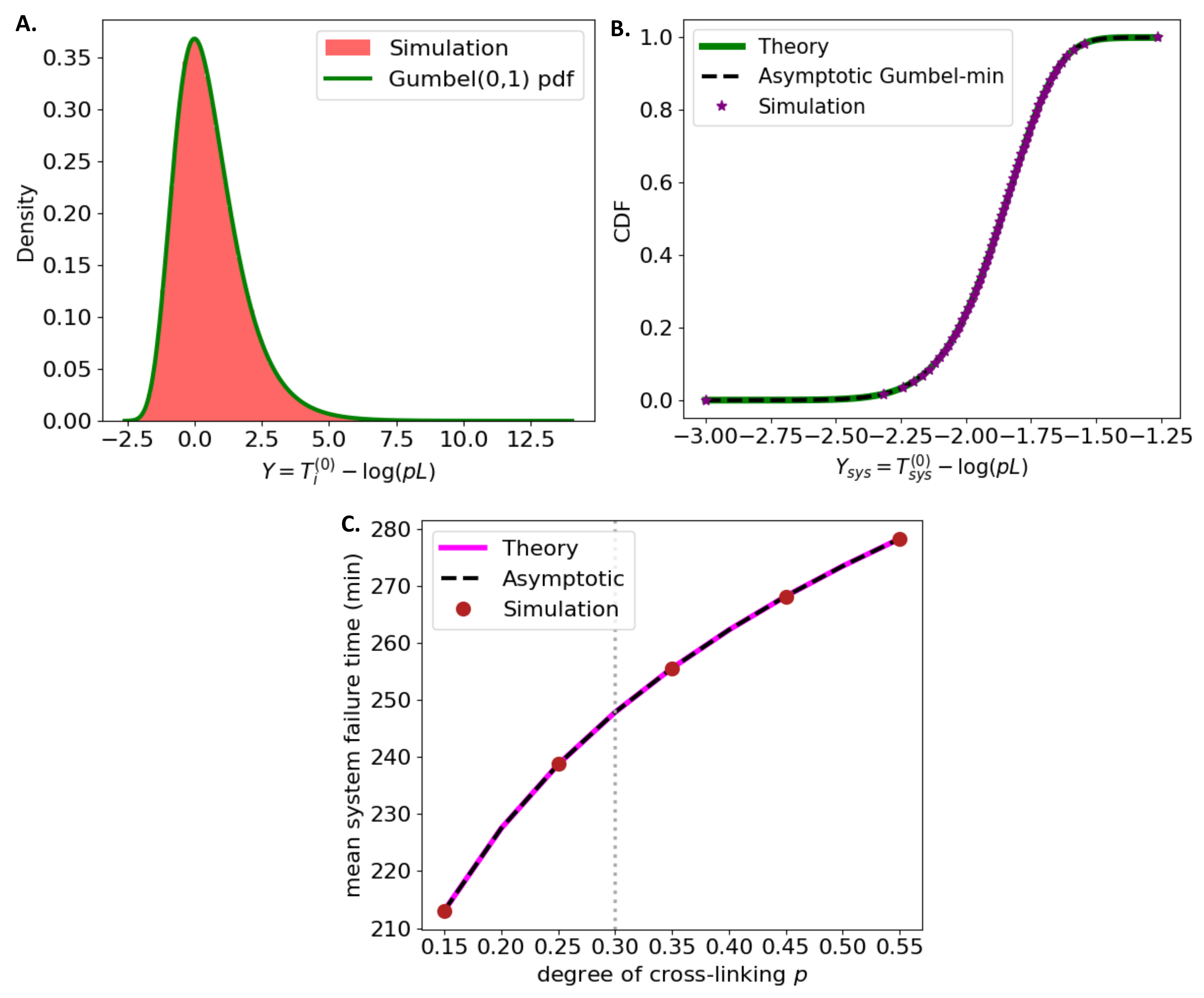}
     \caption{\textbf{Failure of the PG mesh under hydrolytic degradation without mechanical load.} Results are shown for physiologically estimated parameters ($L=3050$, $N_V=444$, $p=0.30$, $r_0=3.32\times10^{-4}\,\mathrm{s^{-1}}$). \textbf{(A)} Distribution of the centered single-cross-section failure time $Y=T_i^{(0)}-\log(pL)$, compared with the standard Gumbel distribution. \textbf{(B)} Empirical CDF of the centered system failure time $Y_{\mathrm{sys}}=T_{\mathrm{sys}}^{(0)}-\log(pL)$, compared with finite-$L$ theory, the asymptotic distribution of the minimum of $N_V$ independent standard Gumbel variables and stochastic simulations. \textbf{(C)} Mean system failure time versus cross-linking degree $p$, with $N_V$ and $L$ fixed, compared with finite $L$ theory, the large-$L$ limit prediction, and stochastic simulations. The dotted line marks the physiological value $p=0.30$. $n=30000$ independent system realizations were used for each value of $p$.}
    \label{fig:figure2}
\end{figure}

\subsubsection*{Asymptotic behaviour: large cross-linking and long cell limits}
We now consider two complementary limits of the no-load model.
First, we study $L\to\infty$ with $N_V$ fixed, to understand
how increasing the number of possible peptide bonds in each
cross-section affects survival. For fixed cross-linking degree
$p>0$, the centered failure time of a single cross-section
converges in distribution to a standard Gumbel random variable:
\begin{equation}
T_i^{(0)}-\log(pL)
\Rightarrow G,
\end{equation}
where $\Rightarrow$ denotes convergence in distribution
(see Appendix A).

At the PG mesh level, failure occurs with the first occurence of a cross-section losing all its peptide bonds. Consequently, the centered system failure time converges to
$M_{N_V}=\min_{1\leq i\leq N_V}G_i$, where the $G_i$ are
independent standard Gumbel random variables (see Appendix A). Its limiting
cumulative distribution function is
\begin{equation}\label{eq:min-gumbel}
F_{M_{N_V}}(y)
=
1-\left(1-e^{-e^{-y}}\right)^{N_V}.
\end{equation}
Fig. \ref{fig:figure2}(A, B) compare this prediction with the finite-$L$ theory and stochastic simulations, showing excellent agreement. This confirms that system failure is governed by the left tail of the single-column failure distribution, because the system fails at the first locally depleted cross-section.

The large-$L$ result also clarifies the role of the cross-linking degree $p$, which determines the limiting system failure time with the centering $\log(pL)$. But as noted, increasing $p$ increases the expected mesh lifetime, but only logarithmically. Thus, additional cross-linking provides progressively diminishing gains in protection. This is highlighted in Fig.\ref{fig:figure2}(C), where the mean system failure time increases with $p$, with stochastic simulations agreeing closely with the finite-$L$ theory and the large-$L$ asymptotic prediction.

We next ask how system size affects failure, thus we study the long-cell limit, $N_V\to\infty$ with local cross-sectional architecture fixed. If the initial condition allows empty cross-sections, then a sufficiently long cell is overwhelmingly likely to contain a failed or critically weak cross-section. This is because each cross-section has a
nonzero probability $(1-p)^L$ of being initially empty
when $p<1$, thus the probability of at least one initially
empty cross-section approaches one as $N_V\to\infty$.
To study time-dependent failure, we therefore condition
on an initially viable mesh in which every cross-section
has at least $m_*\geq1$ peptide bonds. Let $\pi_{m_*}>0$
denote the probability that a cross-section initially
contains exactly $m_*$ bonds.

At early times, failure is dominated by cross-sections
with the smallest initial number of bonds, $m_*$. Thus we get characteristic failure-time scale in dimensionless time as
\begin{equation}\label{eq:failuretime-limit-longcell}
t_{\mathrm{char}}^{(0)}
=
\left(\frac{1}{N_V\pi_{m_*}}\right)^{1/m_*}.
\end{equation}
and the corresponding physical timescale is
$t_{\mathrm{char}}^{(0)}/r_0$ (see Appendix A for details). The dependence on $N_V$
and $m_*$ is verified numerically in
Fig.~\ref{fig:appendix_fig1}.

Thus, even when every cross-section is initially nonempty,
increasing the number of cross-sections reduces the time
to first local failure. A longer mesh contains more
potential weakest-link sites and is therefore more
vulnerable to an unusually early local collapse.
Importantly, this effect does not require heterogeneity:
even when every cross-section initially contains the
same number $k$ of bonds, the characteristic failure
time decreases as $N_V^{-1/k}$.

Together, these results show that the no-load model has two distinct regimes. Increasing the number of bonds per cross-section improves local survival only logarithmically, while increasing the number of cross-sections produces a weakest-link vulnerability.

\subsection*{Failure under hydrolytic action with load}

We next incorporate the mechanical load generated by turgor pressure. The geometric description and failure criterion are as in the no-loading case, with the glycan strands taken to remain intact, while the peptide bonds connecting them are progressively removed by hydrolytic action. The system fails when all peptide bonds are lost in at least one cross-section.

The vertical cross-sections are treated as parallel load-bearing elements, each carrying the same fixed total load $F$, due to turgor pressure (Fig. \ref{fig:schematic}). Within a cross-section, this load is shared equally among all surviving peptide bonds. Hydrolases are again assumed to be uniformly available across the PG mesh, which sets the no-load cleavage rate, but the load borne by surviving bonds modulate rupture rates now.

\begin{figure}[t]
\centering
    \includegraphics[width=\columnwidth]{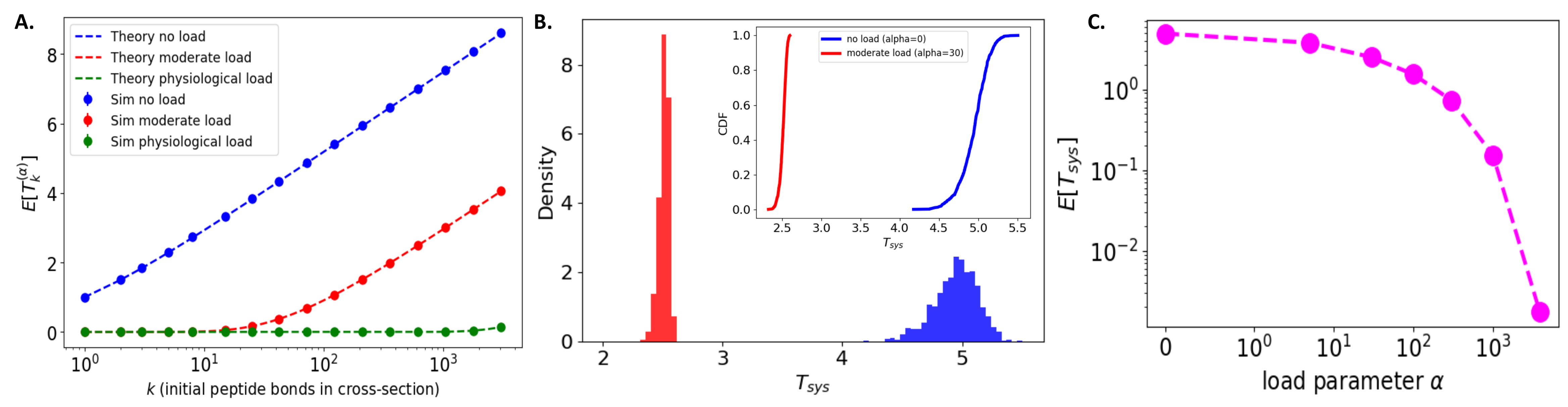}
     \caption{\textbf{Load-dependent failure times.}
(A) Mean single-cross-section
failure time $\mathbb{E}[T_k^{(\alpha)}]$ as a function of the initial
peptide-bond number $k$ for $\alpha=0$ (no load), $\alpha=30$ (moderate load), and
$\alpha=3882$ (physiological load). Dashed lines show the theoretical prediction from
Eq.~\ref{eq:failure-average-cross-section-load} and markers show simulation means. The simulation
results agree closely with the theory. At physiological load, increasing
the initial bond number has little effect on the failure time until
$k$ becomes comparable to $\alpha$, consistent with the
force-dominated regime. (B) Whole-mesh failure-time distributions $T_{\mathrm{sys}}$ for the cases $\alpha=0$ and $\alpha=30$, with CDFs shown
in the inset. (C) Mean whole-mesh failure time
$\mathbb{E}[T_{\mathrm{sys}}]$ as a function of the load parameter
$\alpha$, as obtained from simulations. The mean failure time decreases with increasing load, with a
strong decrease at large $\alpha$. The physiological load parameter is
$\alpha_{\mathrm{phys}}\approx3882$. }
    \label{fig:figure3}
\end{figure}

Suppose that cross-section $i$ contains $X_i(t)=k$
intact peptide bonds. Since the total load $F$ is shared
equally among these bonds, the force on each surviving
bond is $f_i(t)=F/k$. We use a Bell-type force-dependent rupture law \cite{Bell1978,EvansRitchie1997},
\begin{equation}
r(f)=r_0\exp\left(\frac{fx^\ddagger}{k_BT}\right),
\end{equation}
where $x^\ddagger$ is the activation distance and
$k_BT$ is the thermal energy. Defining the dimensionless
load parameter $\alpha=\frac{Fx^\ddagger}{k_BT}$, the rupture rate of a single bond in a cross-section
containing $k$ surviving bonds is
\begin{equation}
r_k=r_0\exp\left(\frac{\alpha}{k}\right).
\end{equation}
Since any one of the $k$ surviving bonds may rupture, the cross-sectional rate is
\begin{equation}
\lambda_{\alpha}(k)
=
r_0 k\exp\left(\frac{\alpha}{k}\right),
\qquad
k\geq 1,
\end{equation}
with $\lambda_{\alpha}(0)=0$. The no-load case corresponds to when $\alpha=0$. Thus, the factor by which load increases the transition rate relative to the no-load case is
\begin{equation}
\frac{\lambda_{\alpha}(k)}{\lambda_0(k)}
=
\exp\left(\frac{\alpha}{k}\right).
\end{equation}
As the previous equation shows, the relative rate of loading vis-a-vis the no loading rate grows as $k$ decreases. Further, in the no-load regime, the total cutting rate of a cross-section, $\lambda_0(k)=r_0k$, decreases monotonically as bonds are removed. On the other hand, in the loading case, the rate exhibits a crossover behaviour- while it may initially decrease as the number of available bonds falls as in the no loading case, but once the cross-section enters the depleted regime \(k<\alpha\), force amplification dominates with the total cutting rate rising sharply, and accelerating towards failure. In the particular case of rod shaped bacteria like \textit{E.coli}, taking turgor pressure $P\sim 1\,\mathrm{atm}$, a cell radius $R\sim 0.5\,\mu\mathrm{m}$, $x^\ddagger\sim 0.2\,\mathrm{nm}$ and $k_{\mathrm B}T\sim4.1\,\mathrm{pN\,nm}$, the load parameter is $\alpha=\frac{Fx^\ddagger}{k_{\mathrm B}T}=\frac{P\pi R^2x^\ddagger}{k_{\mathrm B}T}
\approx 3900$. Thus, for typical cross-section initially containing approximately $k_0\sim 10^3$ bonds, we already have $k_0<\alpha$ and so, the cross-section lies in the force-dominated regime from the beginning, so the total cutting rate increases from the start as bonds are excised, in sharp contrast to the no-loading regime. 
We note that $x^\ddagger=0.2\,\mathrm{nm}$ is
an illustrative molecular-scale activation distance
within the phenomenological Bell framework, and not measured specifically for PG hydrolases.
Our model is formulated in terms of the dimensionless
parameter $\alpha=Fx^\ddagger/(k_BT)$, and its
dependence on this parameter is explored over a
broad range in Fig.~\ref{fig:figure3}(C).
Thus, in particular, the analysis is not restricted to the
particular choice $x^\ddagger=0.2\,\mathrm{nm}$.

We now model the rupture dynamics of peptides in a cross-section which begins with $k$ peptide bonds and evolves as a pure-death continuous-time Markov chain, with transition from $m$ to $m-1$ occurring at rate $\lambda_{\alpha}(m)
=
r_0m\exp\left(\frac{\alpha}{m}\right)$. Its failure time is thus $T_k^{(\alpha)}
=
\sum_{m=1}^{k}\tau_m, \tau_m
\sim
\operatorname{Exp}\left(
r_0m\exp\left(\frac{\alpha}{m}\right)
\right)$ (see Appendix B for details). The expected failure time is thus given by
\begin{equation}\label{eq:failure-average-cross-section-load}
\mathbb{E}\!\left[T_k^{(\alpha)}\right]
=
\frac{1}{r_0}
\sum_{m=1}^{k}
\frac{\exp(-\alpha/m)}{m}.
\end{equation}
Hence,
\begin{equation}
\mathbb{E}\!\left[T_k^{(0)}\right]
-
\mathbb{E}\!\left[T_k^{(\alpha)}\right]
=
\frac{1}{r_0}
\sum_{m=1}^{k}
\frac{1-\exp(-\alpha/m)}{m}
>
0
\end{equation}
which gives the difference between expected time in no load and loading case, illustrating that loading reduces the expected lifetime of every cross-section. 

At the mesh level, the initial numbers of peptide bonds are taken to be
\begin{equation}
K_i=X_i(0)\sim\operatorname{Binomial}(L,p),
\qquad i=1,\ldots,N_V,
\end{equation}
and each cross-section evolves according to the loaded pure-death process described above. The PG mesh fails when the first cross-section becomes completely depleted, so that
\begin{equation}
T_{\mathrm{sys}}^{(\alpha)}
=
\min_{1\leq i\leq N_V}T_i^{(\alpha)}.
\end{equation}
Thus, mechanical loading acts locally by accelerating rupture within depleted cross-sections, while system failure is determined globally by the earliest of the resulting cross-sectional failure times.

We first examine this finite-system model numerically before considering its limiting behavior (the vectorized simulation method used below is compared vis-a-vis full event-driven Gillespie simulation in Fig.~\ref{fig:appendix2}, see Appendix \ref{app:simulations} for details). The simulations compare loaded and unloaded systems generated with the same values of $L$, $N_V$, and $p$, and quantify how the load parameter $\alpha$ changes both single-cross-section and whole-mesh failure times. The effect of mechanical loading on failure times is summarized in Fig.~\ref{fig:figure3}. At the single-cross-section level, Fig.~\ref{fig:figure3}(A) shows $\mathbb{E}[T_k^{(\alpha)}]$ as a function of the initial number of peptide bonds $k$ for $\alpha=0$, $30$, and $3882$, with simulations agreeing closely with Eq.~\ref{eq:failure-average-cross-section-load}. In particular, at physiological loading, increasing the initial bond number provides little additional protection until $k$ becomes comparable to $\alpha$, reflecting the force-dominated regime. In Fig. \ref{fig:figure3}(B) we compare the whole-mesh failure-time distributions in the unloaded case, $\alpha=0$, and for an illustrative loaded case, $\alpha=30$, showing a clear shift toward earlier failure under load. This comparison is extended in Fig. \ref{fig:figure3}(C) over a wider range of $\alpha$ and shows that the mean system failure time, $\mathbb{E}[T_{\mathrm{sys}}]$, decreases monotonically with increasing load, with particularly strong reduction as $\alpha$ increases and approaches the physiological regime, $\alpha_{\mathrm{phys}}\approx3882$. 

\begin{figure}
\centering
    \includegraphics[width= 6 in]{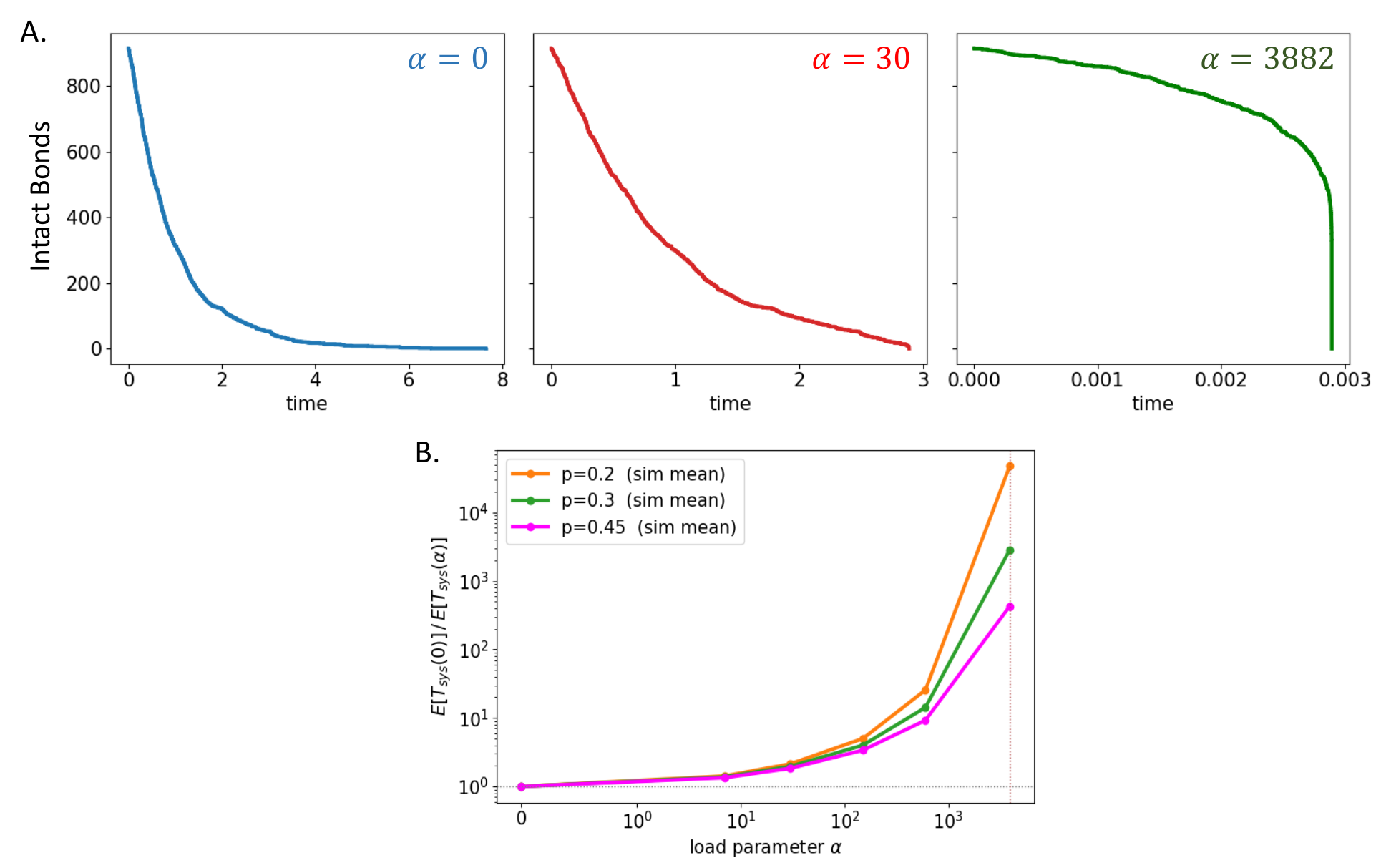}
     \caption{\textbf{Effect of mechanical load on mesh failure.}
(A) Representative single-cross-section bond-loss trajectories,
paired using common random numbers, for the cases $\alpha=0$ (blue),
$\alpha=30$ (red), and physiological load $\alpha=3882$ (green). Increasing load accelerates bond-loss
trajectory in time. At physiological load, most bond loss occurs during
a short terminal collapse, in contrast to the gradual decline at no
load. (B) Lifetime reduction factor,
$\mathbb{E}[T_{\mathrm{sys}}(0)]/\mathbb{E}[T_{\mathrm{sys}}(\alpha)]$, as obtained from simulation means,
as a function of $\alpha$ for three cross-linking degrees $p$. The
physiological estimate $\alpha_{\mathrm{phys}}\approx3882$ is marked by
the vertical dotted line, while the horizontal dotted line indicates
no reduction. At $\alpha_{\mathrm{phys}}$, the lifetime reduction is
approximately $4.7\times10^4$, $2.8\times10^3$, and $4.2\times10^2$ for
$p=0.20$, $0.30$, and $0.45$, respectively
}
    \label{fig:figure4}
\end{figure}

The mechanism underlying this load-induced reduction in failure time is illustrated in Fig.~\ref{fig:figure4}. In Fig. \ref{fig:figure4}(A), we show paired single-cross-section bond-loss trajectories for $\alpha=0$, $30$, and the physiological value $\alpha_{\mathrm{phys}}\approx3882$, using the same underlying random realization in each case. While bond loss is gradual without load, increasing $\alpha$ compresses the trajectory in time, and at physiological load the cross-section undergoes a rapid terminal collapse as force becomes concentrated on the remaining bonds. At the system level, Fig.~\ref{fig:figure4}(B) quantifies this acceleration through the lifetime-reduction factor $\mathbb{E}[T_{\mathrm{sys}}(0)]/\mathbb{E}[T_{\mathrm{sys}}(\alpha)]$. The reduction grows strongly with $\alpha$ and is most pronounced at lower cross-linking degrees, showing how mechanical loading and local cross-link density jointly determine the vulnerability of the PG mesh.

\subsubsection*{Asymptotic behaviour under mechanical load: large cross-linking and long cell limits}

We now consider the two complementary limits of the model under
mechanical load. First, we study $L\to\infty$ with $N_V$
and the load parameter $\alpha$ fixed. For a cross-section
initially containing $k$ bonds, the expected failure time
is asymptotically given by
\begin{equation}
\mathbb{E}\!\left[T_k^{(\alpha)}\right]
\approx
\frac{1}{r_0}
\left[
\log k+\gamma+C(\alpha)
\right].
\end{equation}
where $C(\alpha)<0$ for $\alpha>0$ is the
correction term which depends on the load (see Appendix B). Thus, the leading dependence on the initial number of bonds remains
logarithmic, and increasing cross-linking provides only
a slowly increasing safety margin. Mechanical loading
does not change this leading logarithmic dependence,
but reduces the mean lifetime and accelerates the
final stages of collapse.

However, mechanical loading changes the fluctuations
around this logarithmic behaviour. With initial bond
numbers $K_i\sim\operatorname{Binomial}(L,p)$, we obtain
\begin{equation}
r_0T_i^{(\alpha)}-\log(pL)
\Rightarrow G_\alpha,
\end{equation}
where $G_\alpha$ is a load-dependent limiting random
variable. When $\alpha=0$, $G_0$ follows the standard
Gumbel distribution, whereas for $\alpha>0$, the
limiting distribution is no longer standard Gumbel.
Thus, for fixed $\alpha$, the cross-linking degree $p$
enters through the centering term $\log(pL)$, while
the shape of the centered limiting distribution is
determined by $\alpha$ (see Appendix B for details).

At the mesh level, failure occurs when the first
cross-section becomes completely depleted. Hence,
for fixed $N_V$, we have
$r_0T_{\mathrm{sys}}^{(\alpha)}-\log(pL)
\Rightarrow M_{N_V}^{(\alpha)}$,
where $M_{N_V}^{(\alpha)}$ is the minimum of $N_V$
independent copies of $G_\alpha$. This is the loaded
analogue of the minimum-of-Gumbel result obtained
in the no-load regime (see Eq.~\ref{eq:min-gumbel}).

We next consider the long-cell limit, $N_V\to\infty$,
with $L$ and $\alpha$ fixed. As in the no-load model, initial condition with fixed $L$ and $p<1$ permits initially empty cross-sections in the limit $N_V\to\infty$, since load does not alter the initial configuration. To study genuine time-dependent failure, we therefore condition on an initially viable mesh in which every cross-section contains at least $m_*\geq 1$ bonds. Let $\pi_m$ denote the fraction of cross-sections that begin with exactly $m$ bonds, with $\pi_{m_*}>0$. At early times, failure is dominated by these weakest cross-sections.
Mechanical loading leaves the leading small-time
power unchanged but amplifies their failure
probability by the factor $\exp(\alpha H_{m_*})$,
where $H_{m_*}$ is the corresponding harmonic number.

In this case, after rescaling by
$r_0[N_V\pi_{m_*}\exp(\alpha H_{m_*})]^{1/m_*}$,
the system failure time converges in distribution
to a Weibull random variable $W_{m_*}$, with
survival function
$\mathbb{P}(W_{m_*}>z)=\exp(-z^{m_*})$ (see Appendix B).
The characteristic failure time of the mesh scales as
\begin{equation}
T_{\mathrm{sys}}^{(\alpha)}
\sim
\frac{1}{r_0}
\exp\left(
-\frac{\alpha H_{m_*}}{m_*}
\right)
\left(
\frac{1}{N_V\pi_{m_*}}
\right)^{1/m_*}.
\label{eq:loaded-long-cell-scale}
\end{equation}

Comparison with the no-load result in
Eq.~\ref{eq:failuretime-limit-longcell}
shows that mechanical loading does not change
the weakest-link exponent $1/m_*$. Instead,
it reduces the characteristic failure-time
scale by the multiplicative factor
$\exp(-\alpha H_{m_*}/m_*)<1$ for $\alpha>0$.
Increasing $N_V$ and increasing $\alpha$ therefore
shorten the lifetime through distinct mechanisms:
increasing $N_V$ creates more opportunities for an anomalously early local failure, whereas mechanical loading accelerates the progression of a locally depleted cross-section toward complete collapse.

Importantly, initial heterogeneity is not required for this weakest-link effect. If every cross-section begins with the same number $k_0$ of bonds, Equation~\eqref{eq:loaded-long-cell-scale} then reduces to
\begin{equation}
T_{\mathrm{sys}}^{(\alpha)}
\sim
\frac{1}{r_0}
\exp\left(
-\frac{\alpha H_{k_0}}{k_0}
\right)
N_V^{-1/k_0}.
\label{eq:loaded-homogeneous-long-cell}
\end{equation}
Thus, even an initially homogeneous mesh exhibits earlier first failure as the number of cross-sections increases. Structural heterogeneity is therefore not required for the weakest-link effect, although it can strengthen the effect by introducing cross-sections with smaller initial bond numbers.

Importantly, the weakest-link behaviour is not restricted
to complete depletion of a circumferential cross-section.
The same framework can be applied to localized critical
failure regions containing a smaller number $n_c$ of peptide cross-links, for instance $10-20$,
whose collective removal is assumed to produce a critical
local defect. For $N_{\mathrm{loc}}$ such independent local regions, with an effective local loading parameter $\alpha_{\mathrm{loc}}$, the characteristic failure time scales as

\begin{equation}
T_{\mathrm{char,loc}}^{(\alpha_{\mathrm{loc}})}
\sim
\frac{1}{r_0}
\exp\left(
-\frac{\alpha_{\mathrm{loc}}H_{n_c}}{n_c}
\right)
N_{\mathrm{loc}}^{-1/n_c}.
\end{equation}

Thus, the Weibull limiting form and the weakest-link
mechanism persist for localized failure, although the
scaling exponent and characteristic timescale depend
on the local failure criterion (see Appendix B).

\subsubsection*{Division as mechanical protection}

The previous result suggests a potential mechanical effect of division in mitigating risk of failure that increases as cells elongate uncontrollably. Indeed, cells as they grow, add further possible sites for an anomalously early failure, while load sharing between peptide bonds accelerates the final depletion of any local section that begins to weaken. An elongating cell therefore pays penalty from having more possible failure sites as well as a load-amplification penalty once damage becomes localized due to local depletion of bonds.

By partitioning the cell before $N_V$ becomes too large, division reduces the
risk of weakest-link failure experienced by each individual cell
(the decrease of system lifetime with $N_V$ is illustrated in
Fig.~\ref{fig:appendix_fig1}. Specifically, division lowers the risk of cell failure due to defects in the cell envelope by compartmentalizing failure prone regions, so that a defect arising in one daughter cell does not determine the fate of other daughter cell, potentially continuing lineage spread. This identifies a direct mechanical consequence of
limiting cell length through cell division, as a risk mitigation mechanism. While this does not imply that
division evolved primarily as a mechanical protection mechanism, it shows that
cell-size control can contribute to envelope integrity in addition to its
reproductive function.

More broadly, division or segmentation can provide mechanical protection in growing biological structures by limiting the spatial domain over which rare local defects can arise and trigger failure. From a materials-design perspective, the same principle underpins modular and compartmentalized architectures in the design of tough materials, in which local damage is confined so that failure of one region or module is prevented from propagating into catastrophic failure of the whole system \cite{GordonStrongMaterials}. 

Further, this can be shown more generally and does not require blocked growth- suppose that, over some fixed biologically relevant interval, each
effectively independent cross-section in the mesh has a nonzero probability $q$, with
$0<q<1$, of becoming critically weak. For instance, such a condition holds whenever
local remodelling dynamics remain stochastic and there exists a possible
sequence of degradation and repair events that can drive a local region to a
critical state. If a cell contains $N_{\mathrm{eff}}$ effectively independent
cross-sections, then
\begin{equation}
\mathbb{P}\!\left(
\text{at least one critical cross-section}
\right)
=
1-(1-q)^{N_{\mathrm{eff}}}.
\end{equation}
For fixed $q>0$, this probability increases with $N_{\mathrm{eff}}$ and
approaches unity as $N_{\mathrm{eff}}\to\infty$. Thus, the same weakest-link
principle will apply more generally to a growing bacterial cell provided that
local failure retains a nonzero stochastic probability. Bond insertion and repair may
reduce $q$, but increasing cell length still increases the number of
opportunities for rare local failure. By limiting the number of such
opportunities within an individual cell, division can therefore provide a
mechanically protective effect.

\subsection*{Failure Geometry: Localized Versus Spread-Out Collapse}
We analyzed failure-time distributions above, determining \textit{when} the PG mesh fails, thus the temporal aspects of PG failure. We now endeavor to understand the spatial aspects of PG failure. Specifically, we now explore how much of the PG mesh is degraded by the time of failure and the heterogeneity in terms of damage across the PG mesh at failure. This distinguishes failure caused by widespread bond loss from failure initiated by the local collapse of a single weak cross-section.

We address this question first for finite systems using stochastic simulations. For each initial PG configuration, we simulate both the no-load and load-activated dynamics, allowing the geometry of failure to be compared under the same values of $N_V$, $L$, and $p$, and from similar initial bond configuration.

\subsubsection*{Total bond loss at failure}

Let $B_{\mathrm{fail}}$ denote the total number of peptide bonds broken up to and including the rupture event that causes the first cross-section to fail. We define the fraction of all possible peptide bonds removed before failure by
\begin{equation}
\phi_{\mathrm{fail}}
=
\frac{B_{\mathrm{fail}}}{N_VL}
\end{equation}
so, a small value of $\phi_{\mathrm{fail}}$ indicates that failure occurs after limited global degradation of the PG mesh, consistent with localized collapse while a larger value indicates that bond loss is more broadly distributed before the first cross-section becomes fully depleted. Specifically, a value approaching the
initial cross-linking degree $p$ indicates that most initially present bonds
have been removed before failure.

\subsubsection*{Spatial distribution of damage}

The total number of broken bonds is a relatively coarse metric to determine whether damage is concentrated in one region or distributed across many cross-sections. We therefore define a more refined metric to calculate the damage accumulated in cross-section $i$ at system failure as
\begin{equation}
D_i
=
K_i-X_i\!\left(T_{\mathrm{sys}}\right).
\end{equation}
Clearly, $\sum_{i=1}^{N_V}D_i=B_{\mathrm{fail}}$. Now, to quantify the spatial spread of damage, we use the normalized participation ratio
\begin{equation}
\mathcal{P}_{\mathrm{fail}}
=
\frac{
\left(\displaystyle\sum_{i=1}^{N_V}D_i\right)^2
}{
N_V\displaystyle\sum_{i=1}^{N_V}D_i^2
}
\end{equation}
so, if damage is distributed approximately uniformly across the mesh, then $\mathcal{P}_{\mathrm{fail}}
\approx
1$ and if damage is concentrated in only a small number of cross-sections, then $\mathcal{P}_{\mathrm{fail}}<<1$, reaching a value of order $1/N_V$ when essentially all damage is confined to one cross-section. 

\begin{figure}
\centering
    \includegraphics[width= 6 in]{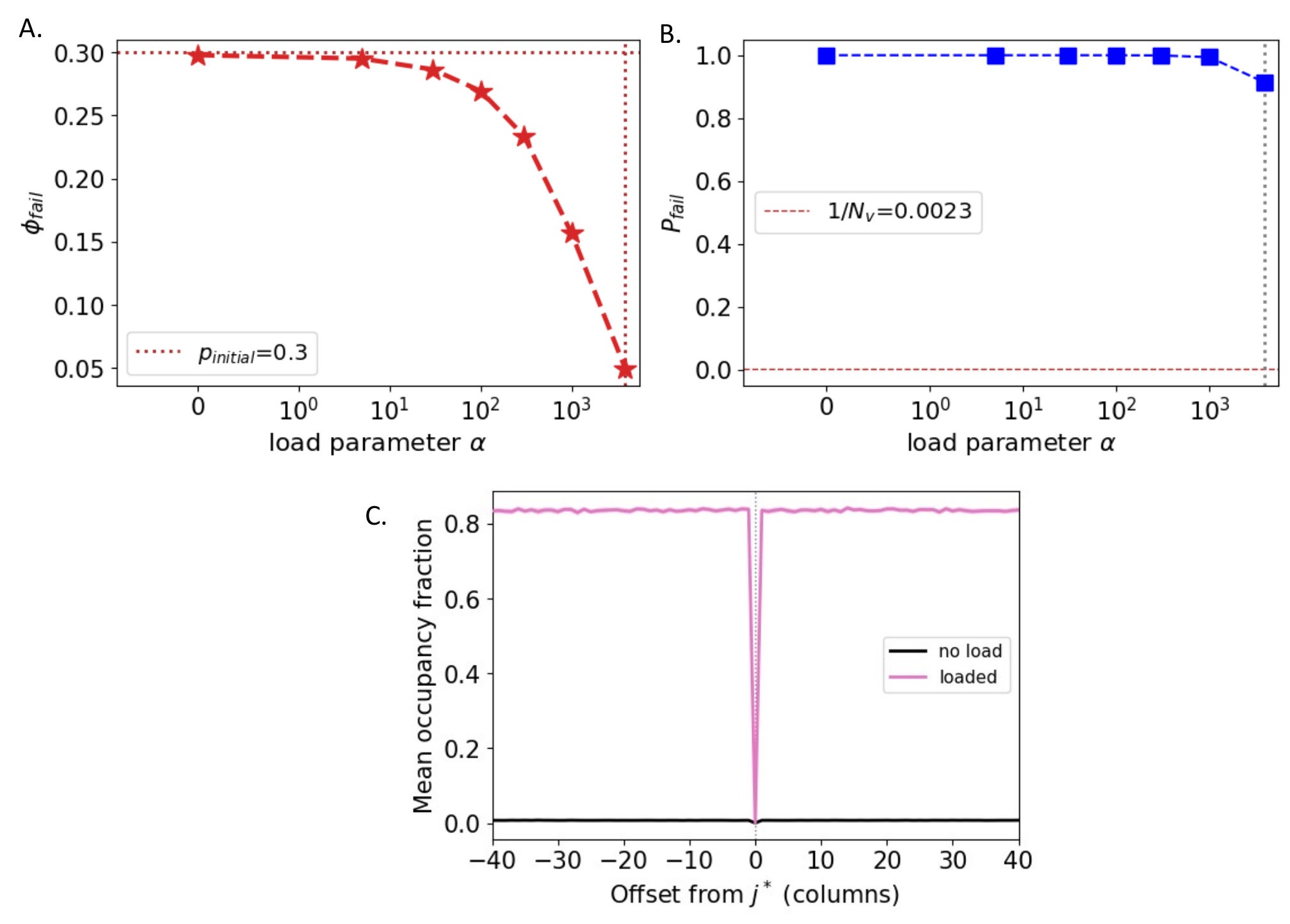}
    \caption{\textbf{Mechanical load shifts mesh failure from global depletion to local collapse.}
(A) Fraction of peptide bonds broken by the time of system failure normalized by number of possible cross-linking sites,
$\phi_{\mathrm{fail}}=B_{\mathrm{fail}}/N_VL$, as a function of the load parameter $\alpha$
($L=3050$, $N_V=444$, $p=0.30$). The initial cross-linking degree
$p_{\mathrm{initial}}=0.3$ and the physiological load parameter
$\alpha_{\mathrm{phys}}\approx3882$ are shown by the horizontal and
vertical dotted lines, respectively. $\phi_{\mathrm{fail}}$ decreases
from $\approx p_{\mathrm{initial}}$ at no load to $\approx0.05$ at
physiological load. (B) Participation ratio $P_{\mathrm{fail}}$ as a
function of $\alpha$. A value of $1/N_V$ would correspond to damage
confined to a single cross-section. $P_{\mathrm{fail}}$ remains close
to 1 over most of the range and decreases to $\approx0.91$ at
physiological load. (C) Mean fraction of initially present bonds remaining at system failure (or mean occupancy fraction at failure),
$X_j(T_{\mathrm{sys}})/K_j$, of the cross-sections,
aligned such that the first-failing cross-section $j^*$ has zero occupancy ratio. Without load, the occupancy away from $j^*$ is $\approx0.01$,
whereas it remains $\approx0.83$ under physiological load. Thus,
without load the mesh is largely depleted when failure occurs, while
under physiological load, local collapse occurs at the first-failing cross-section while the surrounding mesh retains most of its initially present bonds.
}

    \label{fig:figure5}
\end{figure}

\subsubsection*{Effect of load on damage localization}

The expected difference between the two regimes follows from their state-dependent rupture rates. In the no-load regime, conditional on the state $\mathbf{k}
=
(k_1,\dots,k_{N_V})$,
the probability that the next rupture occurs in cross-section $i$ is 
\begin{equation}
    \mathbb{P}_0\!\left(
\text{next rupture occurs in }i
\,\middle|\,
\mathbf{k}
\right)
=
\frac{k_i}{
\displaystyle\sum_{\ell=1}^{N_V}k_\ell
}\end{equation} 
and so, a cross-section that has already lost many bonds therefore becomes less likely to receive the next rupture event. Under load, the corresponding probability is
\begin{equation}
\mathbb{P}_{\alpha}\!\left(
\text{next rupture occurs in }i
\,\middle|\,
\mathbf{k}
\right)
=
\frac{
k_i\exp(\alpha/k_i)
}{
\displaystyle
\sum_{\ell=1}^{N_V}
k_\ell\exp(\alpha/k_\ell)
}.
\end{equation}

For two cross-sections with $k_i<k_j$, their relative odds satisfy
\begin{equation}
\frac{
\mathbb{P}_{\alpha}\!\left(
\text{next rupture occurs in }i
\,\middle|\,
\mathbf{k}
\right)
}{
\mathbb{P}_{\alpha}\!\left(
\text{next rupture occurs in }j
\,\middle|\,
\mathbf{k}
\right)
}
=
\frac{k_i}{k_j}
\exp\left[
\alpha
\left(
\frac{1}{k_i}
-
\frac{1}{k_j}
\right)
\right].
\end{equation}
The exponential factor is greater than one. Hence, compared with the no-load dynamics, load increases the relative probability that rupture returns to the more depleted cross-section.

This calculation provides a mechanism for enhanced localization under load. In the no-load regime, damage tends to be distributed among cross-sections that still contain many available bonds. Under load, an emerging weak cross-section becomes increasingly susceptible to further rupture and may proceed rapidly to complete depletion, though it does not by itself prove that load always produces smaller $B_{\mathrm{fail}}$ in every finite system. To understand this, we therefore quantify the effect numerically through $\phi_{\mathrm{fail}}$ and $\mathcal{P}_{\mathrm{fail}}$, as functions of the load parameter $\alpha$, the initial cross-linking degree $p$, and the system size $N_V$, as summarized
in Fig.~\ref{fig:figure5}. Figure~\ref{fig:figure5}(A) shows that the fraction of all possible cross-linking sites
removed by the time of system failure, $\phi_{\rm fail}$, decreases markedly
with increasing $\alpha$: in the no-load regime it is close to the initial
cross-linking degree $p$, whereas at physiological load, it decreases dramatically to $\sim 0.05$.
Thus, mechanical loading permits system failure after substantially less global
degradation of the PG mesh. Figure~\ref{fig:figure5}(B) shows that the participation ratio
$P_{\rm fail}$ also decreases with load, indicating that accumulated damage
becomes less uniformly distributed across the mesh, however with
$P_{\rm fail}\approx0.91$ even at physiological load, indicating that bond loss prior to
failure remains spread over many cross-sections rather than being confined to
the failing one. This distinction is clearly revealed in Fig.~\ref{fig:figure5}(C), where
cross-sections are aligned relative to the first-failing cross-section $j^*$.
Without load, the surrounding cross-sections are themselves almost completely
depleted when $j^*$ reaches zero, whereas under physiological loading they
retain approximately $83\%$ of their initially present bonds. Thus, mechanical
loading changes the character of failure from one associated with near-global
depletion to the terminal collapse of a locally weakened cross-section while
most of the surrounding mesh remains substantially intact.
\newpage

\subsubsection*{Long-cell limit}
For fixed $L$ and fixed load parameter, under the initially
viable-mesh assumption with minimum initial bond number
$m_*\geq1$, the mean failure time
$\mathbb{E}\!\left[T_{\mathrm{sys}}\right]$ scales as
$N_V^{-1/m_*}$. Since the total rupture rate of the mesh
is at most of order $N_V$, it follows that
$\mathbb{E}\!\left[B_{\mathrm{fail}}\right]$ scales as $N_V^{1-1/m_*}$. Consequently, we have that
$\frac{B_{\mathrm{fail}}}{N_VL}\rightarrow 0$ as $N_V\to\infty$
(see Appendix~\ref{app:failuregeometry} for details).
Moreover, if $A_{\mathrm{fail}}$ denotes the number of
cross-sections experiencing at least one rupture before
system failure, then $A_{\mathrm{fail}}\leq B_{\mathrm{fail}}$,
implying $\frac{A_{\mathrm{fail}}}{N_V}\rightarrow 0$. So, we have that in the long-cell limit, damage is confined to a vanishing fraction of cross-sections. In other words, in the long cell limit, damage is highly localized since asymptotically almost all cross-sections remain completely untouched when the first one fails. 

The corresponding finite-size dependence is shown in
Fig.~A.3. Over the range of $N_V$ considered,
$\phi_{\rm fail}$ depends only weakly on system size
without load, whereas under mechanical loading it
decreases more appreciably as $N_V$ increases, with
the strongest dependence at physiological load.
These results are consistent with the long-cell
prediction of an increasingly localized failure.

\subsection*{Role of Cross-Linking Degree}

The degree of cross-linking $p$ determines the fraction of available peptide-bonding sites that are initially occupied. As  noted, $K_i=X_i(0)\sim\operatorname{Binomial}(L,p)$ and the cross-section contains on average $\mathbb{E}[K_i]=pL$ peptide bonds. Increasing $p$ therefore increases the average number of bonds present in the PG mesh and under mechanical load, it has an additional effect as a larger number of surviving bonds also reduces the force carried by each individual bond.

We first examine the effect of cross-linking directly in finite systems by stochastic simulation, varying $p$ while holding $L$ and $N_V$ fixed. We perform this comparison in both the no-load and load-activated regimes. For each value of $p$, we measure the distribution and mean of the system failure time,
\begin{equation}
T_{\mathrm{sys}}
=
\min_{1\leq i\leq N_V}T_i,
\end{equation}
and quantify the effect of increasing cross-linking on delaying mesh failure.

\subsubsection*{Cross-linking in the PG mesh}

In the no-load regime, increasing $p$ increases the number of bonds that must be removed before a cross-section becomes depleted. It also lowers the probability of initially weak cross-sections, since the lower tail of $K_i\sim\operatorname{Binomial}(L,p)$ moves toward larger bond numbers as $p$ increases. Both effects increase the time to first failure.

For fixed $L$, the benefit of increasing cross-linking however must eventually saturate because each cross-section can contain at most $L$ peptide bonds. As $p\to1$, the initial distribution concentrates at $K_i=L$, and thus, in the no-load case, $\mathbb{E}\!\left[T_i^{(0)}\right]
\rightarrow
\frac{\sum_{m=1}^L 1/m}{r_0}$, which is the maximum mean lifetime attainable by increasing $p$ at fixed $L$. 

Under load, cross-linking provides an additional layer of protection. As noted, cross-section containing $k$ surviving bonds has total rupture rate $\lambda_{\alpha}(k)
=
r_0k
\exp\left(\frac{\alpha}{k}\right)$ and so, the force-dependent amplification experienced by an individual bond is $\exp\left(\frac{\alpha}{k}\right)$. Increasing $p$ raises the typical initial value of $k$ and hence decreases the initial force carried by each bond. More highly cross-linked cross-sections consequently begin farther from force-amplified depleted states.

We therefore examine how the cross-linking degree $p$ affects system survival
and failure geometry. For the no-loading case, dependence of the mean system failure time
on $p$ was studied in Fig.\ref{fig:figure2}(C), while the interaction between cross-linking and
mechanical loading is quantified through the lifetime-reduction factor in
Fig.~\ref{fig:figure4}(B). The geometry-at-failure is also explored by studying $\phi_{\mathrm{fail}}$ and $\mathcal{P}_{\mathrm{fail}}$, as introduced in the previous section, across the same range of $p$ to determine whether increasing cross-linking changes not only the time to failure but also the degree to which damage is localized at failure. The effect of cross-linking degree on the geometry of failure is shown in
Fig.~\ref{fig:appendix4}. Figure~\ref{fig:appendix4}(A) shows that $\phi_{\rm fail}$ increases with $p$ in both
the no-load and physiological-load regimes, although it remains substantially
smaller under load. Thus, more highly cross-linked meshes can sustain a larger
amount of total bond loss before the first cross-section fails, while mechanical
loading continues to promote failure at a substantially smaller level of global
degradation. Figure~\ref{fig:appendix4}(B) shows a distinct effect on the spatial distribution of
damage. In the no-load regime, $P_{\rm fail}$ remains close to unity over the
range of $p$ considered, whereas under physiological loading it increases
strongly with $p$. Hence, at low cross-linking degree, load produces relatively
concentrated damage at failure, while increasing $p$ makes the accumulated
damage progressively more broadly distributed across the mesh.

These results highlight two distinct ways in which cross-linking
protects the PG mesh. First, increasing the cross-linking degree
increases the number of bonds in a typical cross-section, thereby
increasing its lifetime and reducing the force carried by individual
bonds under mechanical loading. However, as shown by the large-$L$
analysis, the improvement in lifetime is only logarithmic, so
additional cross-linking provides diminishing returns. Second,
increasing cross-linking reduces the abundance of weakly connected
cross-sections that can initiate early failure. As established by
the long-cell limit, the abundance of these weakest regions
determines the characteristic failure-time scale, while the
weakest-link exponent remains unchanged when the minimum initial
bond number $m_*$ is fixed. Thus, mesh reliability depends not
only on the average degree of cross-linking but also on the
distribution of bonds across cross-sections. Two meshes with
similar average cross-link density can therefore have very different
failure times if their distributions of weakly connected regions are
different.

\subsection*{Hydrolase activity and protection against collapse}

We now ask by how much hydrolase activity must be reduced to protect
the PG mesh under mechanical loading. We consider two criteria:
survival with a prescribed probability over a given time interval,
and recovery of the mean lifetime of the unloaded system. Let $r_0$ be the per-bond hydrolytic cutting rate at a fixed reference activity level, and $0<a\leq 1$ be the relative activity parameter, where $a=1$ denotes the reference activity and $a<1$ denotes hydrolase downregulation. We model changes in hydrolase activity as a uniform
rescaling of the cleavage rate, so $a$ represents
the fraction of the reference hydrolytic activity that remains, so that
$r_0\to ar_0$. Here we assume that, the cleavage hazard is proportional to the effective
concentration or activity of active enzyme and that
hydrolase activity and mechanical activation act independently. So, under load, when a cross-section contains $k$ surviving peptide bonds, the per-bond cutting rate becomes
\begin{equation}
r_{a,\alpha}(k)
=
a r_0
\exp\left(\frac{\alpha}{k}\right),
\end{equation}
and the total transition rate of the cross-section is $\lambda_{a,\alpha}(k)
=
a r_0 k
\exp\left(\frac{\alpha}{k}\right)$. The activity factor $a$ multiplies every rupture rate by the same amount. Consequently, reducing hydrolase activity does not change which cross-section is most likely to undergo the next rupture. At current state
$\mathbf{k}=(k_1,\dots,k_{N_V})$, the probability that the next rupture occurs in cross-section $i$ is
\begin{equation}
\mathbb{P}_{a,\alpha}
\left(
\text{next rupture occurs in }i
\,\middle|\,
\mathbf{k}
\right)
=
\frac{
k_i\exp(\alpha/k_i)
}{
\displaystyle
\sum_{j=1}^{N_V}
k_j\exp(\alpha/k_j)
},
\end{equation}

\begin{figure}
\centering
    \includegraphics[width=\columnwidth]{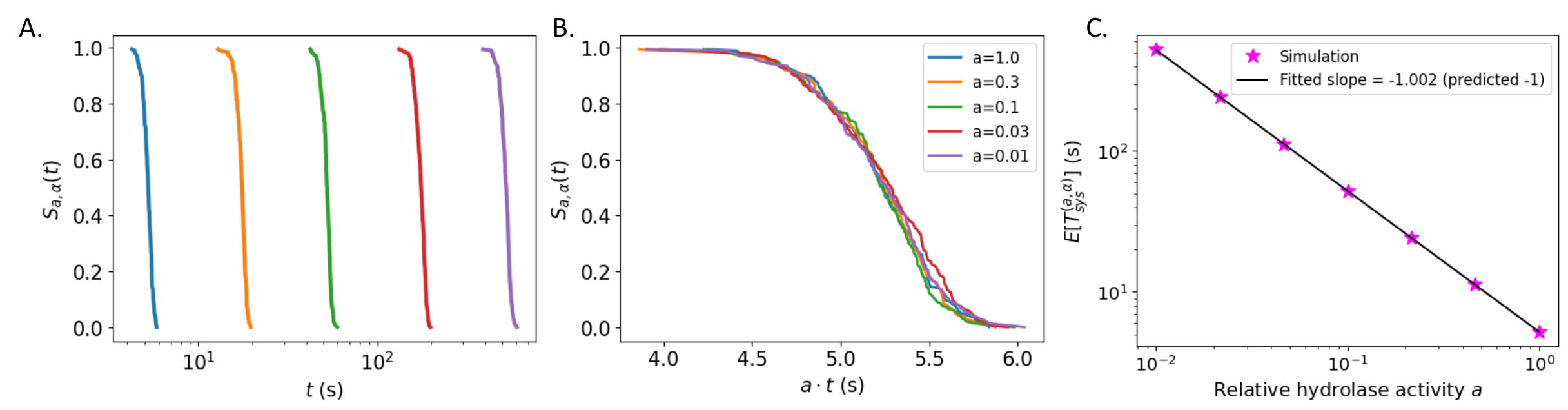}
   \caption{\textbf{Hydrolase activity controls the timescale of mesh failure.}
The physiological load $\alpha=\alpha_{\mathrm{phys}}\approx3882$ is held fixed, with
$L=3050$, $N_V=444$, and $p=0.30$.
\textbf{(A)} System survival probability $S_{a,\alpha}(t)$ for relative hydrolase
activities $a=1,\,0.3,\,0.1,\,0.03,$ and $0.01$. Reducing hydrolase activity shifts
the failure-time distribution to longer times.
\textbf{(B)} The same survival curves plotted against the rescaled time $at$.
Their collapse is consistent with
$S_{a,\alpha}(t)=S_{1,\alpha}(at)$ (The colors in the legend correspond to same activities in (A) and (B)).
\textbf{(C)} Mean system failure time as a function of relative hydrolase activity.
Markers show simulation means and the solid line shows a log--log fit with slope
$-1.002$, in agreement with the exact scaling
$E[T_{\mathrm{sys}}^{(a,\alpha)}]
=a^{-1}E[T_{\mathrm{sys}}^{(1,\alpha)}]$.
Panels (A)--(B) use 200 independent system realizations per value of $a$, and
panel (C) uses 150 realizations per value of $a$.}

    \label{fig:figure6}
\end{figure}

\begin{figure}
\centering
    \includegraphics[width=\columnwidth]{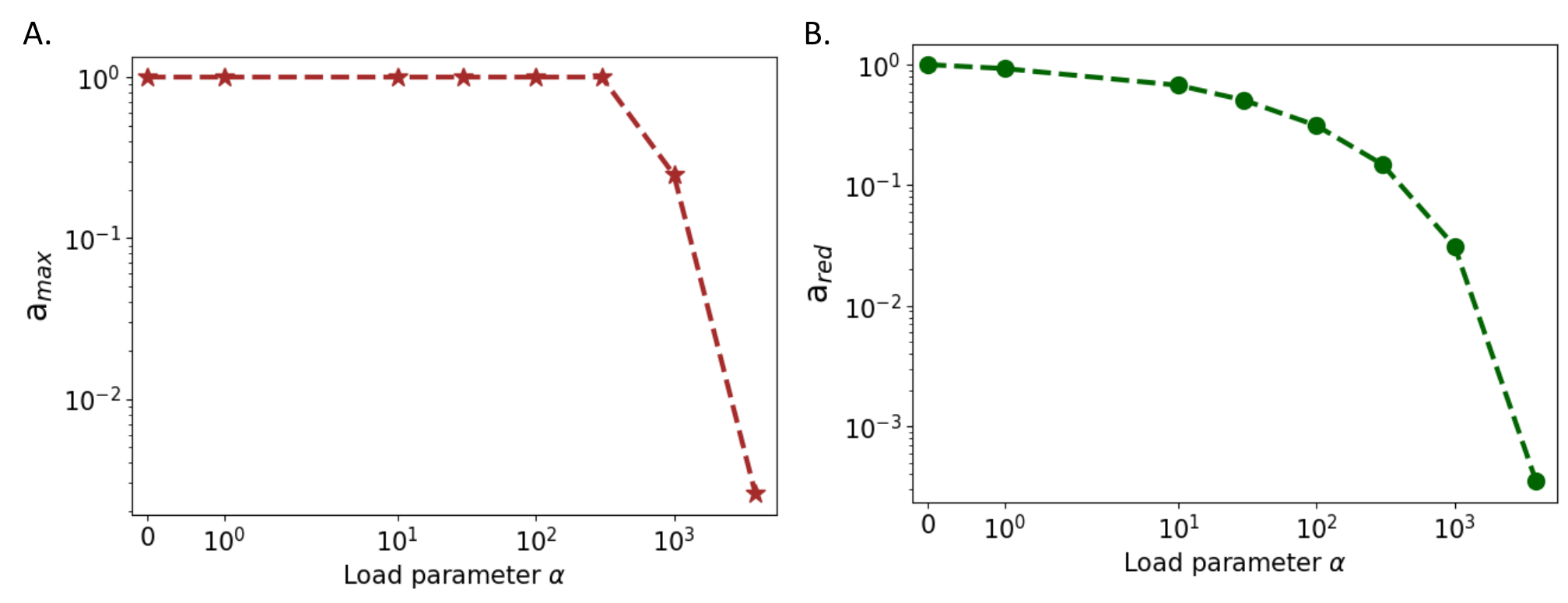}
   \caption{\textbf{Hydrolase reduction can compensate for mechanical load.}
(A) Maximum hydrolase activity $a_{\max}$ that gives at least $95\%$
survival up to $T_{\mathrm{protect}}=30$ min. $a_{\max}$ remains close to
1 at low loads but decreases rapidly at high $\alpha$, reaching
$\approx2.6\times10^{-3}$ at the physiological load
$\alpha\approx3882$. (B) Relative activity reduction factor $a_{\mathrm{red}}$, required
for a loaded mesh to recover the mean failure time of the unloaded mesh.
At $\alpha\approx3882$, this requires
$a_{\mathrm{red}}\approx3.5\times10^{-4}$.}

    \label{fig:figure7}
\end{figure}

which is independent of $a$, indicating that activity modulation leaves the sequence of bond-removal events unchanged and can only influence the waiting times between them. Let $T_{\mathrm{sys}}^{(a,\alpha)}$ denote the system failure time at activity $a$ and load parameter $\alpha$. We then have
\begin{equation}
T_{\mathrm{sys}}^{(a,\alpha)}
\overset{d}{=}
\frac{1}{a}
T_{\mathrm{sys}}^{(1,\alpha)}.
\end{equation}
Therefore,
\begin{equation}
\mathbb{E}
\left[
T_{\mathrm{sys}}^{(a,\alpha)}
\right]
=
\frac{1}{a}
\mathbb{E}
\left[
T_{\mathrm{sys}}^{(1,\alpha)}
\right].
\end{equation}
More generally, if $S_{a,\alpha}(t)
=
\mathbb{P}
\left(
T_{\mathrm{sys}}^{(a,\alpha)}>t
\right)$ denotes the survival probability, then $S_{a,\alpha}(t)
=
S_{1,\alpha}(at)$ (see Appendix D). Thus, lowering hydrolase activity protects the mesh by stretching its entire failure-time distribution. For instance, reducing activity by one half doubles median failure-time, while reducing activity tenfold increases median failure time by a factor of ten. These relations are illustrated in Fig.~\ref{fig:figure6}. Specifically, Fig.\ref{fig:figure6}(A) shows that
reducing the relative hydrolase activity $a$ shifts the entire system
survival curve toward longer times. When time is rescaled by the activity,
however, the survival curves collapse onto one another (Fig.~\ref{fig:figure6}(B)), as
predicted by $S_{a,\alpha}(t)=S_{1,\alpha}(at)$. The corresponding mean
failure times are shown in Fig.~\ref{fig:figure6}(C) and follow the predicted inverse scaling
$\mathbb{E}[T_{\mathrm{sys}}^{(a,\alpha)}]\propto a^{-1}$, illustrating that hydrolase downregulation changes the timescale of failure
without changing the underlying sequence of rupture events.

Motivated by the possibility that reduced hydrolase activity may allow the cell wall to survive for a certain period when PG synthesis is inhibited by antibiotics, we specify a minimum survival time $T_{\mathrm{protect}}$ and an acceptable failure probability $\varepsilon$, requiring
\begin{equation}
\mathbb{P}
\left(
T_{\mathrm{sys}}^{(a,\alpha)}
>
T_{\mathrm{protect}}
\right)
\geq
1-\varepsilon.
\end{equation}
Let $Q_{\varepsilon}^{(\alpha)}$ denote the $\varepsilon$-quantile of the
failure time at reference hydrolase activity, defined by $\mathbb{P}\!\left(
T_{\mathrm{sys}}^{(1,\alpha)}
\leq
Q_{\varepsilon}^{(\alpha)}
\right)
=
\varepsilon$. The protection condition is then satisfied whenever
\begin{equation}
a
\leq
a_{\mathrm{max}}
=
\min
\left\{
1,
\frac{
Q_{\varepsilon}^{(\alpha)}
}{
T_{\mathrm{protect}}
}
\right\}
\end{equation}
giving the highest level of hydrolase activity compatible with the prescribed survival requirement (see Appendix D).

Next we ask how much hydrolase activity must be
reduced for a loaded PG mesh to recover the mean lifetime of
the unloaded system, which gives a direct measure of the
reduction needed to compensate for mechanical loading. This is given by
\begin{equation}
a_{\mathrm{red}}
=
\frac{
\mathbb{E}
\left[
T_{\mathrm{sys}}^{(1,\alpha)}
\right]
}{
\mathbb{E}
\left[
T_{\mathrm{sys}}^{(1,0)}
\right]
}.
\end{equation}
Since load accelerates failure, $a_{\mathrm{red}}<1$. Thus, an activity level $a=a_{\mathrm{red}}$ restores the no-load mean survival time, while $a<a_{\mathrm{red}}$ gives a mean lifetime longer than the no-load reference value. The magnitude of the hydrolase reduction required for protection is shown in
Fig.~\ref{fig:figure7}. In particular, Fig~\ref{fig:figure7}(A) gives the maximum relative hydrolase activity compatible
with at least $95\%$ survival up to the prescribed protection time
$T_{\mathrm{protect}}=30$ min. While little reduction is required at low
loads, the admissible activity decreases sharply as $\alpha$ increases,
reaching approximately $2.6\times10^{-3}$ at physiological load.
Figure~\ref{fig:figure7}(B) illustrates the complementary criterion of restoring the unloaded
mean failure time. The required activity fraction $a_{\mathrm{red}}$ also
decreases strongly with load, reaching approximately
$3.5\times10^{-4}$ at $\alpha_{\mathrm{phys}}$, highlighting the substantial hydrolase downregulation required in the model to compensate for the strong acceleration of failure produced by physiological mechanical loading. Thus, we have quantitative estimates of hydrolase activity reduction under two different protection
criteria: survival over a prescribed interval and compensation
for the reduction in lifetime caused by mechanical loading.

In the long-cell limit, this compensation factor has a simple analytical form. Recall the loaded survival probability $\mathbb{P}
\left(
T_{\mathrm{sys}}^{(a,\alpha)}>t
\right)
\approx
\exp\left[
-N_V\pi_{m_*}
\exp(\alpha H_{m_*})
(a r_0t)^{m_*}
\right]$. Comparing this with the no-load survival law at reference activity shows that the complete asymptotic first-failure distribution is restored when $a_{\mathrm{red}}
=
\exp\left(
-\frac{\alpha H_{m_*}}{m_*}
\right)$ (see Appendix D). Hence, in the long-cell regime, hydrolase downregulation can exactly compensate for the load-induced reduction of the weakest-link failure-time scale. This compensation is temporal rather than mechanical. Indeed, $\frac{
\lambda_{a,\alpha}(k)
}{
\lambda_{a,0}(k)
}
=
\exp\left(\frac{\alpha}{k}\right)$, which is independent of $a$. Lowering activity therefore does not remove the increased susceptibility of depleted cross-sections under load, nor does it change the spatial pathway of rupture but rather, it protects the PG mesh by slowing the progression along that pathway and delaying the time at which local collapse occurs.

\section{Discussion}

We developed a minimal stochastic-mechanical theory for failure of the Gram-negative PG wall when insertion of new material is arrested while hydrolytic cleavage continues. The model separates two distinct sources of vulnerability. At the level of a single cross-section, increasing the number of peptide bonds increases lifetime only logarithmically, so that additional cross-linking provides diminishing returns. In the no-load regime, the centered failure time converges to a Gumbel distribution, whereas mechanical loading  fundamentally alters the centered limiting law and accelerates the collapse of depleted cross-sections. At the level of the whole mesh, increasing the number of cross-sections introduces a weakest-link effect: in the long-cell limit, the failure time follows a Weibull-type scaling controlled by the weakest initially allowed regions, while load decreases the failure-time scale without changing the weakest-link exponent. The weakest-link dependence on the number of cross-sections also suggests a mechanical benefit of limiting cell length through division, by reducing the number of potential sites for anomalously early local failure. Importantly, failure need not be accompanied by extensive global degradation: in the long-cell limit both the fraction of bonds removed and the fraction of cross-sections experiencing any damage vanish at failure. Cross-linking therefore protects the wall in two ways, by increasing the number of bonds available within individual cross-sections and by suppressing rare weak regions. Hydrolase downregulation acts differently, by rescaling the
failure-time distribution, without removing the load-dependent preference for rupture in already
depleted regions.

Several experimental observations are qualitatively consistent with the failure picture obtained from the model. Studies of $\beta$-lactam-treated \textit{E.~coli} have shown that lysis can
begin after only partial loss of the PG layer, with PG loss under different
$\beta$-lactams reported not to exceed approximately $20$--$35\%$ prior to
the onset of lysis, rather than requiring uniform destruction of the sacculus
\cite{vanHeijenoort2011}. Microscopy and single-cell studies further show that inhibition of PG synthesis can produce localized bulges, lesions and spatially structured lysis events \cite{Huang2008,Chung2009,Wong2021}, qualitatively paralleling our result that local failure can occur while much of the mesh remains intact. The predicted protective effect of lowering hydrolytic activity also has an experimental analogue: deletion of multiple PG hydrolases in \textit{E.~coli} retards penicillin-induced lysis \cite{Heidrich2002}, consistent with the general conclusion that continued hydrolytic activity strongly influences the time between inhibition of wall synthesis and mechanical collapse. Finally, experiments showing that turgor and mechanical stress influence bacterial-wall deformation and lysis support the inclusion of mechanics as an essential component of the failure problem \cite{Wong2021, Deng2011}. These observations support the broader picture emerging from the model: bacterial-wall failure can be initiated by local defects, can occur well before complete global PG degradation, and depends jointly on hydrolytic activity and mechanical loading.

Our model has connections with several other stochastic and mechanical
problems in biology. One example is a cluster of parallel cell-adhesion bonds
sharing an external force. Stochastic bond-cluster models show that, in the
weak-rebinding regime, cluster lifetime grows only logarithmically with bond
number and can fall rapidly as the applied load increases
\cite{ErdmannSchwarz2004}. Force can also
affect enzymatic degradation. Tensile force has been shown to accelerate
MMP-1-mediated collagen proteolysis \cite{Adhikari2011}, whereas in other
collagens force can stabilize the substrate against enzymatic
degradation \cite{Flynn2010}. 

Other examples arise in the context of population mode. In a
pure-death population in which individuals die independently at a constant
rate, extinction occurs when the last individual dies, giving a mean
extinction time proportional to
$H_N\sim\log N$. This is precisely the logarithmic dependence obtained for the
lifetime of a cross-section in the no-load case. More general population
models, including reproduction, density dependence, and environmental
fluctuations, can give very different extinction-time scalings
\cite{OvaskainenMeerson2010}. A related role for redundancy occurs in adaptive
immunity, where increasing lymphocyte-receptor diversity lowers the
probability that a pathogen escapes recognition
\cite{DeBoerPerelson1993,PerelsonWeisbuch1997}. In each of these examples,
adding independent bonds, individuals, or recognition elements can improve
robustness, but the increase in lifetime or protection need not be
proportional to the amount of redundancy added.

The present model is a minimal stochastic model and its exact
limiting distributions should therefore be interpreted as
predictions of the idealized model. The PG mesh is coarse-grained
into cross-sections characterized only by the number of surviving
peptide bonds. Initial bond numbers are assumed to be independent,
hydrolases are taken to be uniformly available, and the load on a
cross-section is shared equally among its surviving bonds.
Detailed network connectivity, spatial correlations, and non-local
load redistribution are not included. We also define failure as
complete depletion of a single cross-section and use the Bell-type
rate as a phenomenological description of force-dependent cleavage. We note that a sufficiently large local pore can permit membrane
bulging and subsequent lysis \cite{Huang2008,Wong2021,Daly2011}.
As illustrated in Appendix B, our stochastic framework
can accommodate failure of smaller load-bearing regions,
while retaining a weakest-link character.
This represents a specific assumption of mechanically
assisted cleavage, rather than a universal force-response law for PG hydrolases. Other force-response mechanisms, for instance saturating or force-inhibited hydrolytic activity, are possible and could modify the quantitative predictions of load-induced acceleration. The present analysis therefore examines how mechanically assisted cleavage interacts with the structure and cross-linking of the PG mesh to determine failure dynamics.
Nevertheless, these simplifications allow three mechanisms to
be isolated clearly: diminishing protection from increasing
local bond number, extreme-value vulnerability arising from
increasing the number of possible failure sites, and mechanical
acceleration of already depleted regions. Several extensions
are natural. Allowing neighbouring cross-sections to interact,
for example, would make it possible to study how a local defect
develops into a larger crack. One could also relax the
assumptions of uniform hydrolase activity and equal load sharing
to see how sensitive the results are to spatial heterogeneity.
Experimentally, measurements of local PG cross-linking and of
how hydrolase-mediated cleavage changes under force would help
identify the regime relevant in vivo. Taken together, our model
effectively delineates how local architecture, system size,
mechanical load, and hydrolytic activity jointly determine
PG-wall failure.

\section*{Acknowledgments}
 
G.R. acknowledges support from the DST INSPIRE (DST/INSPIRE/04/2024/003426) and the New Faculty Seed Grant from IIT Delhi (IM00039). Part of the work was done when I.P. was visiting IMJ-PRG, Universit\'e Paris Cit\'e, supported by CNRS IEA grant GAOA. 

\section*{Conflicts of Interest}
The authors declare no conflicts of interest.


\appendix
\section*{Appendix}
\renewcommand{\thefigure}{A.\arabic{figure}}
\setcounter{figure}{0}

\section{Failure under hydrolytic action and no loading}

\subsubsection*{Microscopic description of the failure mode}

We assume that hydrolases are uniformly available across the PG mesh at a fixed local concentration. Consequently, in the absence of mechanical load, every intact peptide bond is cut independently at the same effective hydrolytic rate $r_0$.

Let
\begin{equation}
E(t)
=
\sum_{j=1}^{N_V}X_j(t)
\end{equation}
denote the total number of intact peptide bonds at time $t$. Since each of the $E(t)$ surviving bonds has cutting rate $r_0$, the total cutting rate of the mesh is
\begin{equation}
\Lambda(t)
=
r_0E(t).
\end{equation}

Conditional on the current configuration $\{X_j(t)\}$, the probability that the next cut occurs in cross-section $i$ is
\begin{equation}
\mathbb{P}
\left(
\text{next cut occurs in cross-section }i
\mid
\{X_j(t)\}
\right)
=
\frac{r_0X_i(t)}{r_0E(t)}
=
\frac{X_i(t)}{E(t)}.
\end{equation}

The total cutting rate of the mesh is $\Lambda(t)=r_0E(t)$.
The cutting rate in cross-section $i$ is therefore obtained
by multiplying the total rate by the probability that the
next cut occurs in that cross-section, as given in the
previous equation. Thus,
\begin{equation}
\lambda_i^{\mathrm{phys}}(t)
=
\Lambda(t)\frac{X_i(t)}{E(t)}
=
r_0X_i(t).
\end{equation}

Thus, a cross-section with more intact peptide bonds has a larger total cutting rate, because each surviving bond is exposed to the same local hydrolase concentration.

For convenience, we introduce the dimensionless time
\begin{equation}
s
=
r_0t_{\mathrm{phys}}.
\end{equation}
In this dimensionless time, every intact peptide bond has unit cutting rate. All failure times below are expressed in this dimensionless time unless stated otherwise. Physical failure times are recovered by dividing the corresponding dimensionless times by $r_0$.

\subsection*{Dynamics in a single cross-section}
Let
\begin{equation}
K_i
=
X_i(0)
\end{equation}
be the initial number of peptide bonds in cross-section $i$. 
Conditional on $K_i=k$, the $k$ bond lifetimes in dimensionless
time are independent exponential random variables with unit
rate. Since the cross-section fails when the last bond is
removed, its failure time is
\begin{equation}
T_i^{(0)}
=
\max\{\xi_1,\ldots,\xi_k\},
\qquad
\xi_j\sim\operatorname{Exp}(1).
\end{equation}
Alternatively, we can model the successive bond-removal events - when $m$ bonds remain, each has unit cutting rate, so the
total rate of the next cleavage event is $m$. The exponential distribution has the memoryless property, so we have that the successive
waiting times are independent, so
\begin{equation}
T_i^{(0)}
=
\sum_{m=1}^{k}\tau_m,
\qquad
\tau_m\sim\operatorname{Exp}(m).
\end{equation}

Since an exponential random variable with rate $m$ has mean
$1/m$, taking expectations gives
\begin{equation}
\mathbb{E}\left[T_i^{(0)}\mid K_i=k\right]
=
\sum_{m=1}^{k}\mathbb{E}[\tau_m]
=
\sum_{m=1}^{k}\frac{1}{m}
=
H_k
=
\log k+\gamma+O(k^{-1}).
\end{equation}
where $H_k$ is the $k$-th harmonic number and $\gamma$ is Euler's constant. In physical time,
\begin{equation}
\mathbb{E}
\left[
T_{i,\mathrm{phys}}^{(0)}
\mid
K_i=k
\right]
=
\frac{H_k}{r_0}.
\end{equation}

Conditional on $K_i=k$, the failure-time distribution is
\begin{equation}
\mathbb{P}
\left(
T_i^{(0)}\leq t
\mid
K_i=k
\right)
=
\left(1-e^{-t}\right)^k.
\end{equation}
since the cutting events are  independent and follow exponential distribution. Now, replacing $t$ by
$t+\log k$ and taking limit, we get
\begin{equation}
\mathbb{P}
\left(
T_i^{(0)}-\log k\leq t
\mid
K_i=k
\right)
=
\left(
1-\frac{e^{-t}}{k}
\right)^k
\longrightarrow
e^{-e^{-t}}
\end{equation}
as $k\to\infty$. Thus, conditional on $K_i=k$, we have that $T_i^{(0)}-\log k$ converges in distribution to the Gumbel distribution, expressed as
\begin{equation}
T_i^{(0)}-\log k
\Rightarrow
G,
\end{equation}
where $G$ is a standard Gumbel random variable, with the result holding conditional to the initial number of bonds, $K_i$, being equal to $k$.

We now remove the conditioning on the initial bond number to derive the general convergence result. Since $K_i \sim \operatorname{Binomial}(L,p)$,
the unconditional failure-time distribution is
\begin{equation}
\begin{aligned}
\mathbb{P}(T_i^{(0)}\leq s)
&=
\sum_{k=0}^{L}
\binom{L}{k}p^k(1-p)^{L-k}(1-e^{-s})^k\\
&=
\sum_{k=0}^{L}
\binom{L}{k}
[p(1-e^{-s})]^k(1-p)^{L-k}\\
&=
(1-pe^{-s})^L
\end{aligned}
\end{equation}
where the last equality follows from the binomial theorem. To recenter, we let $s=t+\log(pL)$, and thus
\begin{equation}
\begin{aligned}
\mathbb{P}\left(
T_i^{(0)}-\log(pL)\leq t
\right)
=
\left[
1-p\exp\left(-t-\log(pL)\right)
\right]^L\
=
\left(
1-\frac{e^{-t}}{L}
\right)^L\
\longrightarrow
e^{-e^{-t}},
\end{aligned}
\end{equation}
for any $t\in\mathbb{R}$ as $L\to\infty$. So, we have 
\begin{equation}
T_i^{(0)}-\log(pL)
\Rightarrow
G.
\end{equation}

\subsection*{Global failure}

The PG mesh is defined to fail when the first cross-section loses all of its peptide bonds. Thus,
\begin{equation}
T_{\mathrm{sys}}^{(0)}
=
\min_{1\leq i\leq N_V}
T_i^{(0)}.
\end{equation}

For fixed $N_V$, as $L\to\infty$,
\begin{equation}
T_{\mathrm{sys}}^{(0)}
-
\log(pL)
\Rightarrow
M_{N_V},
\end{equation}
where $M_{N_V}
=
\min_{1\leq i\leq N_V}G_i$ and $G_1,\dots,G_{N_V}$ are independent standard Gumbel random variables. It then follows that the cumulative distribution function of $M_{N_V}$ is
\begin{equation}
\mathbb{P}(M_{N_V}\leq y)
=
1-
\left(
1-e^{-e^{-y}}
\right)^{N_V}.
\end{equation}

In physical time, the corresponding result is
\begin{equation}
r_0T_{\mathrm{sys,phys}}^{(0)}
-
\log(pL)
\Rightarrow
M_{N_V}.
\end{equation}

We next study the no-load regime in the limit $N_V\to\infty$, with  $L$ fixed. This corresponds to a cell becoming longer and longer while its local cross-sectional architecture remains fixed.

Let $K_i
=
X_i(0),
K_i\sim\operatorname{Binomial}(L,p)$, independently across cross-sections. If $p<1$, then a given cross-section has probability $\mathbb{P}(K_i=0)
=
(1-p)^L$ of being empty at time zero. Therefore,
\begin{equation}
\mathbb{P}
\left(
\text{at least one initially empty cross-section}
\right)
=
1-
\left(
1-(1-p)^L
\right)^{N_V}.
\end{equation}
Since $L$ is fixed and $(1-p)^L>0$,
\begin{equation}
1-
\left(
1-(1-p)^L
\right)^{N_V}
\longrightarrow
1
\end{equation}
as $N_V\to\infty$.

Thus, under the strict failure rule that one empty cross-section implies system failure, a sufficiently long cell is overwhelmingly likely to contain an initially failed cross-section.

To study time-dependent failure rather than failure already present in the initial condition, we condition on an initially viable mesh. We assume that every cross-section has at least $m_*\geq1$ peptide bonds at time zero and the number of peptide bonds in the cross-sections follow some distribution. So, let
\begin{equation}
\pi_m
=
\mathbb{P}(K_i=m)
\end{equation}
denote the initial bond-number distribution, conditional on $m_*\leq m\leq L$, where we also take $\pi_{m_*}>0$. If a cross-section has $m$ bonds to start with, then as seen previously, 
\begin{equation}
F_m^{(0)}(t)
=
\mathbb{P}
\left(
\text{Cross-section failure time}\leq t
\right)
=
\left(1-e^{-t}\right)^m.
\end{equation}
It follows that for small $t$, we have $F_m^{(0)}(t)
\sim
t^m$. The failure distribution of a randomly selected cross-section is therefore
\begin{equation}
F_{\mathrm{col}}^{(0)}(t)
=
\sum_{m=m_*}^{L}
\pi_mF_m^{(0)}(t).
\end{equation}
At early times, the contribution from the minimum initial bond number dominates, giving
\begin{equation}
F_{\mathrm{col}}^{(0)}(t)
\sim
\pi_{m_*}t^{m_*}
\end{equation}
Since the system survives beyond time $t$ only if every cross-section does as well, we get $
\mathbb{P}
\left(
T_{\mathrm{sys}}^{(0)}>t
\right)
=
\left(
1-F_{\mathrm{col}}^{(0)}(t)
\right)^{N_V}$, and so using the early-time scaling, we have
\begin{equation}
\mathbb{P}
\left(
T_{\mathrm{sys}}^{(0)}>t
\right)
\sim
\exp
\left[
-N_V\pi_{m_*}t^{m_*}
\right].
\end{equation}

So, the characteristic failure transition time is given by $N_V\pi_{m_*}t^{m_*}\sim 1$. Therefore, in dimensionless time,
\begin{equation}
T_{\mathrm{sys}}^{(0)}
\sim
\left(
\frac{1}{N_V\pi_{m_*}}
\right)^{1/m_*}.
\end{equation}
In physical time,
\begin{equation}
T_{\mathrm{sys,phys}}^{(0)}
\sim
\frac{1}{r_0}
\left(
\frac{1}{N_V\pi_{m_*}}
\right)^{1/m_*}.
\end{equation}

\section{Rupture dynamics under hydrolytic action and with loading}

Consider $N_v$ vertical cross-sections, each consisting of two glycan strands connected by peptide bonds. The glycan strands are assumed to remain intact, while peptide bonds rupture under the combined action of hydrolases and mechanical load. Peptide rupture is assumed to be irreversible on the timescale considered, and rebinding is therefore neglected.

For each cross-section $i=1,\ldots,N_v$, let $X_i(t)\in\{0,1,\ldots,L\}$ denote the number of intact peptide bonds at time $t$. Initially, $X_i(0)=K_i, K_i\sim\mathrm{Binomial}(L,p)$, independently across cross-sections. Each cross-section is assumed to carry a constant force $F$, shared equally among its surviving peptide bonds. Thus, when $X_i(t)=m>0$, the force carried by each intact peptide is $f=\frac{F}{m}$. Using the Bell-type force-dependent rupture rate $r(f)=r_0\exp\left(\frac{f x^\ddagger}{k_{B}T}\right)$, the rupture rate of an individual peptide is $r_0\exp\left(\frac{Fx^\ddagger}{m k_{B}T}\right)$, so with $m$ intact peptides, the total rupture rate of the cross-section is $\lambda_\alpha(m)
=
r_0 m \exp\left(\frac{\alpha}{m}\right)$ where $\alpha=\frac{Fx^\ddagger}{k_{B}T}$. Hence, each cross-section evolves as a pure-death continuous-time Markov chain with transitions $m\longrightarrow m-1$ occurring at rate $\lambda_\alpha(m)$, with $\lambda_\alpha(0)=0$. 

Conditional on the cross-section initially containing $k$ intact peptide bonds, the pure-death process goes through the states
\[
k,k-1,\ldots,1,0
\]
sequentially, since each rupture removes exactly one peptide bond and rebinding is neglected. When $m$ bonds remain, the total rupture rate is $\lambda_\alpha(m)$, so the waiting time until the next rupture is exponentially distributed with parameter $\lambda_\alpha(m)$. Therefore, the total time required for the cross-section to reach the terminating state $0$ is the sum of the waiting times in each intermediate state, given as
\begin{equation}\label{eq:cross-section-loaded}
T_k^{(\alpha)}
=
\sum_{m=1}^{k}\tau_m,
\end{equation}
where the waiting times $\tau_m\sim \operatorname{Exp}(
r_0 m \exp(\frac{\alpha}{m}))$ are independent. Therefore,
\begin{equation}
\mathbb{E}\left[T_k^{(\alpha)}\right]
=
\frac{1}{r_0}
\sum_{m=1}^{k}
\frac{\exp(-\alpha/m)}{m}.
\end{equation}

In the absence of load, $\alpha=0$, giving $\mathbb{E}\left[T_k^{(0)}\right]
=
\frac{1}{r_0}
\sum_{m=1}^{k}\frac{1}{m}$. Thus,
\begin{equation}
\mathbb{E}\left[T_k^{(0)}\right]
-
\mathbb{E}\left[T_k^{(\alpha)}\right]
=
\frac{1}{r_0}
\sum_{m=1}^{k}
\frac{1-\exp(-\alpha/m)}{m}
>0
\end{equation}
for $\alpha>0$. Assuming the cross-sections evolve independently, the system fails when the first cross-section loses all of its peptide bonds. Hence,
\begin{equation}
T_{\mathrm{sys}}
=
\min_{1\leq i\leq N_v}
T_{K_i}^{(\alpha)}.    
\end{equation}

\subsection*{Large-$L$ limit under fixed load}

Throughout this subsection, the load parameter $\alpha$ is held fixed as
$L\to\infty$. Conditional on a cross-section initially containing $k$
bonds, we have as above, $\mathbb{E}\!\left[T_k^{(\alpha)}\right]
=
\frac{1}{r_0}
\sum_{m=1}^{k}
\frac{e^{-\alpha/m}}{m}$. Now,
\[
\frac{e^{-\alpha/m}-1}{m}
\sim
-\frac{\alpha}{m^2},
\]
as $m\to\infty$. Therefore, $C(\alpha) = \sum_{m=1}^{\infty} \frac{e^{-\alpha/m}-1}{m}$ converges absolutely and in fact, $C(\alpha)<0$ for every $\alpha>0$. Note now that
\[
\sum_{m=1}^{k}\frac{e^{-\alpha/m}}{m}
=
H_k+
\sum_{m=1}^{k}\frac{e^{-\alpha/m}-1}{m}.
\]
For fixed $\alpha$, the second sum converges to
$C(\alpha)$. Since
$H_k
\approx
\log k+\gamma$
it follows that
\begin{equation}
\mathbb{E}\!\left[T_k^{(\alpha)}\right]
\approx
\frac{1}{r_0}
\left[
\log k+\gamma+C(\alpha)
\right].
\end{equation}

Let $E_m$ be independent unit-rate exponential
random variables, $E_m\sim\operatorname{Exp}(1)$.
Since the waiting time $\tau_m$ has exponential
rate $r_0m e^{\alpha/m}$, it can be represented as $\tau_m=\frac{E_m}{r_0m e^{\alpha/m}}$. Consequently, Eq.~\ref{eq:cross-section-loaded}
gives
\begin{equation}
r_0T_k^{(\alpha)}-\log k
=
H_k-\log k
+
\sum_{m=1}^{k}
\frac{e^{-\alpha/m}E_m-1}{m}.
\end{equation}
and we also rewrite
\begin{equation}
\frac{e^{-\alpha/m}E_m-1}{m}
=
\frac{e^{-\alpha/m}(E_m-1)}{m}
+
\frac{e^{-\alpha/m}-1}{m}.
\end{equation}
As shown above, $C(\alpha)$ converges, so the right part, when summed, converges absolutely. As for the random terms, note for any $m$, $
\frac{e^{-\alpha/m}(E_m-1)}{m}$ has mean zero and also
\begin{equation}
\operatorname{Var}\!\left[
\frac{e^{-\alpha/m}(E_m-1)}{m}
\right]
=
\frac{e^{-2\alpha/m}}{m^2}
\leq
\frac{1}{m^2}.
\end{equation}
Thus, $\sum_{m=1}^{\infty}
\frac{e^{-\alpha/m}(E_m-1)}{m}$ converges almost surely by Kolmogorov's two-series theorem and it follows that $\sum_{m=1}^{\infty}
\frac{e^{-\alpha/m}E_m-1}{m}$ also converges almost surely.
Therefore,
\begin{equation}\label{eq:loaded_gumbel}
r_0T_k^{(\alpha)}-\log k
\Rightarrow
G_\alpha,
\end{equation}
where $G_\alpha
=
\gamma+
\sum_{m=1}^{\infty}
\frac{e^{-\alpha/m}E_m-1}{m}$. For $\alpha=0$, $G_0
=
\gamma+
\sum_{m=1}^{\infty}
\frac{E_m-1}{m}$, which is the standard Gumbel random variable. 

Note, for $\alpha>0$, $\mathbb{E}[G_\alpha]
=
\gamma+C(\alpha)
<
\gamma$,
whereas standard Gumbel random variable has mean $\gamma$.
Thus $G_\alpha$ is not the standard Gumbel law when $\alpha>0$.
Similarly, the variance $\operatorname{Var}(G_\alpha)
=
\sum_{m=1}^{\infty}
\frac{e^{-2\alpha/m}}{m^2}
<
\frac{\pi^2}{6}$, which is again different from standard Gumbel distribution.

We now allow the initial bond number to be random. Let $K_i\sim\operatorname{Binomial}(L,p), 0<p<1$. By weak law of large numbers, we have $K_i/pL\rightarrow 1$ in probability and so we have
$\log K_i-\log(pL)
=
\log\!\left(\frac{K_i}{pL}\right)
\longrightarrow
0$
in probability. Since $K_i\to\infty$ in probability, it follows from Eq. \ref{eq:loaded_gumbel} that $r_0T_i^{(\alpha)}-\log K_i
\Rightarrow
G_\alpha$. Therefore,
\begin{equation}
r_0T_i^{(\alpha)}-\log(pL)
=
\left[
r_0T_i^{(\alpha)}-\log K_i
\right]
+
\left[
\log K_i-\log(pL)
\right],
\end{equation}
and since the first term on the right converges in distribution and the last term converges in probability, Slutsky's theorem gives
\begin{equation}
r_0T_i^{(\alpha)}-\log(pL)
\Rightarrow
G_\alpha.
\end{equation}

Finally, consider PG mesh containing a fixed number $N_V$ of independent
cross-sections. Joint convergence then gives
\begin{equation}
\left(
r_0T_1^{(\alpha)}-\log(pL),
\dots,
r_0T_{N_V}^{(\alpha)}-\log(pL)
\right)
\Rightarrow
\left(
G_{\alpha,1},
\dots,
G_{\alpha,N_V}
\right),
\end{equation}
where the $G_{\alpha,i}$ are independent copies of $G_\alpha$. 

Since the minimum is a continuous function,
the continuous mapping theorem applied to the
preceding joint convergence gives the limiting
distribution of the system failure time as

\begin{equation}
r_0T_{\mathrm{sys}}^{(\alpha)}
-
\log(pL)
\Rightarrow
M_{N_V}^{(\alpha)},
\end{equation}
where $M_{N_V}^{(\alpha)}
=
\min_{1\leq i\leq N_V}G_{\alpha,i}$. If $F_\alpha$ denotes the cumulative distribution function of
$G_\alpha$, then
\begin{equation}
\mathbb{P}\!\left(
M_{N_V}^{(\alpha)}\leq y
\right)
=
1-
\left[
1-F_\alpha(y)
\right]^{N_V}.
\end{equation}

\subsection*{Long-cell limit under load}

Again, for a cross-section with $m$ bonds, the failure time is $T_m^{(\alpha)}
=
\sum_{j=1}^{m}\tau_j, \tau_j\sim
\operatorname{Exp}\left(
r_0 j e^{\alpha/j}
\right)$, with the waiting times independent. Denote $\lambda_j
=
r_0 j e^{\alpha/j}$. 
Since $T_m^{(\alpha)}$ is the sum of $m$ independent
exponential waiting times with rates
$\lambda_1,\ldots,\lambda_m$, we have
\begin{equation}
\mathbb{P}(T_m^{(\alpha)}\leq t)
=
\int_{\substack{s_j\geq0\\
\sum_{j=1}^{m}s_j\leq t}}
\prod_{j=1}^{m}
\lambda_j e^{-\lambda_j s_j}
\,ds_1\cdots ds_m.
\end{equation}
As $t$ becomes smaller and smaller, the exponential factors approach
$1$ uniformly over the integration region, whose
volume is $t^m/m!$. Hence, for small $t$, we have
\begin{equation}
\mathbb{P}(T_m^{(\alpha)}\leq t)
\sim
\frac{\prod_{j=1}^{m}\lambda_j}{m!}t^m.
\end{equation}

Now $\prod_{j=1}^{m}\lambda_j
= r_0^m m!\,e^{\alpha H_m}$,
so, for small $t$, we have
\begin{equation}
\mathbb{P}\!\left(
T_m^{(\alpha)}\leq t
\right)
\sim
e^{\alpha H_m}(r_0t)^m
\end{equation}

Let $\pi_m$ denote the initial bond-number distribution, with $m_*$ the smallest value satisfying $\pi_{m_*}>0$. The failure probability of a randomly chosen cross-section is
\begin{equation}
F_{\mathrm{col}}^{(\alpha)}(t)
=
\sum_{m\geq m_*}
\pi_m
\mathbb{P}\!\left(
T_m^{(\alpha)}\leq t
\right).
\end{equation}

At early times, the smallest power of $t$ dominates since $t^n$ decreases as $n$ increases, so $F_{\mathrm{col}}^{(\alpha)}(t)
\sim
\pi_{m_*}
e^{\alpha H_{m_*}}
(r_0t)^{m_*}$. Hence, for $N_V$ independent cross-sections and for small $t$,
\begin{equation}
\mathbb{P}\!\left(
T_{\mathrm{sys}}^{(\alpha)}>t
\right)
=
\left[
1-F_{\mathrm{col}}^{(\alpha)}(t)
\right]^{N_V}
\approx
\exp\!\left[
-N_V\pi_{m_*}
e^{\alpha H_{m_*}}
(r_0t)^{m_*}
\right]
\end{equation}
using the approximation $1-x \approx e^{-x}$ for small $x$. 

Let $
Z_{N_V}
=
\left[
N_V\pi_{m_*}
e^{\alpha H_{m_*}}
\right]^{1/m_*}
r_0T_{\mathrm{sys}}^{(\alpha)}$. For fixed $z>0$, we then see that $P(Z_{N_V}>z)=P(T_{\mathrm{sys}}^{(\alpha)}>t)$ where $t= 
\frac{z}{
r_0
\left[
N_V\pi_{m_*}
e^{\alpha H_{m_*}}
\right]^{1/m_*}
}$. Since $F_{\mathrm{col}}^{(\alpha)}(t)
\sim
\pi_{m_*}e^{\alpha H_{m_*}}(r_0t)^{m_*}$, and substituting $t=t_{N_V}$, we obtain $N_VF_{\mathrm{col}}^{(\alpha)}(t_{N_V})
\longrightarrow z^{m_*}$. Therefore,
\begin{equation}
\mathbb{P}(Z_{N_V}>z)
\longrightarrow
e^{-z^{m_*}}
\end{equation}
as $N_V\rightarrow \infty$. Thus, $Z_{N_V}
\Rightarrow
W_{m_*}$, where $\mathbb{P}(W_{m_*}>z)
=
e^{-z^{m_*}}$, which is the Weibull-type distribution. The characteristic failure-time scales as
\begin{equation}
T_{\mathrm{sys}}^{(\alpha)}
\sim
\frac{1}{r_0}
\exp\!\left(
-\frac{\alpha H_{m_*}}{m_*}
\right)
\left(
\frac{1}{N_V\pi_{m_*}}
\right)^{1/m_*}.
\end{equation} and in fact, we have
\begin{equation}
\mathbb{E}\!\left[
T_{\mathrm{sys}}^{(\alpha)}
\right]
\sim
\frac{
\Gamma\!\left(1+1/m_*\right)
}{r_0}
\exp\!\left(
-\frac{\alpha H_{m_*}}{m_*}
\right)
\left(
\frac{1}{N_V\pi_{m_*}}
\right)^{1/m_*}.
\end{equation}

\subsection*{Localized critical failure regions}

Our model can also be applied to localized
critical defects, without requiring complete depletion of
an entire circumferential cross-section. 
Consider a spatially localized region containing
$n_c$ initially intact, neighboring
peptide cross-links, whose collective removal will produce a critical gap/opening, leading to cell lysis
(for illustration, $n_c$ may be taken to be of
order $10$-$20$). Let $F_{\mathrm{loc}}$ denote the effective force acting on such a region and define
$\alpha_{\mathrm{loc}}
=F_{\mathrm{loc}}x^\ddagger/(k_BT)$.

Assuming equal load sharing among the surviving bonds
within this region, the total rupture rate when $m$
bonds remain is
$r_0m\exp(\alpha_{\mathrm{loc}}/m)$.
Thus, the local failure time follows the same pure-death
dynamics as before. In particular, its small-time
failure probability satisfies
\begin{equation}
\mathbb{P}(T_{\mathrm{loc}}\leq t)
\sim
\exp(\alpha_{\mathrm{loc}}H_{n_c})
(r_0t)^{n_c}.
\end{equation}

Now consider $N_{\mathrm{loc}}$ candidate localized
critical regions in the PG mesh, and we assume that their failure
times are independent. Let $T_{\mathrm{sys,loc}}$
denote the time to the first failure among these
regions. By the same argument used above,
\begin{equation}
\mathbb{P}(T_{\mathrm{sys,loc}}>t)
\approx
\exp\left[
-N_{\mathrm{loc}}
\exp(\alpha_{\mathrm{loc}}H_{n_c})
(r_0t)^{n_c}
\right]
\end{equation}
on the corresponding early-failure timescale.

Consequently, as $N_{\mathrm{loc}}\to\infty$ with
$n_c$ and $\alpha_{\mathrm{loc}}$ fixed,
\begin{equation}
r_0
\left[
N_{\mathrm{loc}}
\exp(\alpha_{\mathrm{loc}}H_{n_c})
\right]^{1/n_c}
T_{\mathrm{sys,loc}}
\Rightarrow W_{n_c},
\end{equation}
where
$\mathbb{P}(W_{n_c}>z)=\exp(-z^{n_c})$.

Thus, localized critical failure also exhibits
weakest-link behaviour, with the characteristic
failure time decreasing as
$N_{\mathrm{loc}}^{-1/n_c}$. 

More generally, a critical pore may involve
bonds distributed across several neighbouring
cross-sections, and different candidate pore regions
may overlap. In this setting, failure of the entire
mesh occurs when the first such critical region
fails. The underlying weakest-link mechanism remains
applicable. 
Under suitable assumptions, the first-failure
statistics can again exhibit Weibull-type
behaviour for critical regions requiring
a fixed finite number of cleavages. The precise exponent
and characteristic timescale will depend on the
number and arrangement of bonds involved in the
critical failure event. A detailed treatment of
these geometrical features requires an explicit
spatial model of the PG network and will be taken
up in future work.

\section{Asymptotic localization of failure}\label{app:failuregeometry}
\subsection*{Long-cell limit}

Let $N(t)$ denote the total number of rupture events that have occurred by time $t$. Then $B_{\mathrm{fail}}
=
N\!\left(T_{\mathrm{sys}}\right)$. For fixed $L$ and fixed load parameter $\alpha$, define
\begin{equation}
C_{L,\alpha}
=
r_0\max_{1\leq k\leq L}
\left\{
k\exp\!\left(\frac{\alpha}{k}\right)
\right\}.
\end{equation}
This constant is finite because $L$ and $\alpha$ are fixed. If $\Lambda(t)$ denotes the total rupture rate of the mesh at time $t$, then $\Lambda(t)
=
\sum_{i=1}^{N_V}
r_0X_i(t)\exp{\frac{\alpha}{X_i(t)}}
\leq
N_VC_{L,\alpha}$.
But since $\mathbb{E}\!\left[B_{\mathrm{fail}}\right]
=
\mathbb{E}\!\left[
\int_0^{T_{\mathrm{sys}}}\Lambda(t)\,dt
\right]$, using the rate bound we have
\begin{equation}
\mathbb{E}\!\left[B_{\mathrm{fail}}\right]
\leq
N_VC_{L,\alpha}
\mathbb{E}\!\left[T_{\mathrm{sys}}\right].
\end{equation}
The long-cell expected survival time established above gives $\mathbb{E}\!\left[T_{\mathrm{sys}}\right]
\sim
N_V^{-1/m_*}$ (under the initially viable-mesh assumption, with minimum initial bond number $m^*\geq1$) and hence,
\begin{equation}
\mathbb{E}\!\left[B_{\mathrm{fail}}\right]
\sim N_V^{1-1/m_*},
\end{equation}
and therefore
\begin{equation}
\mathbb{E}\!\left[
\frac{B_{\mathrm{fail}}}{N_VL}
\right]
\sim N_V^{-1/m_*}
\longrightarrow
0.
\end{equation}
as $N_V\rightarrow\infty$. Using Markov's inequality, we get
\begin{equation}
\frac{B_{\mathrm{fail}}}{N_VL}
\longrightarrow
0
\end{equation}
in probability.

\section{Scaling with hydrolase activity}
\label{app:activity_scaling}

We derive the activity-scaling results used in the main text. Let $a\in(0,1]$ denote the relative hydrolase activity, and let $r_0$ be the per-bond hydrolytic cutting rate at reference activity.

\subsection*{Single-cross-section scaling}

For a cross-section containing $m$ surviving bonds under load parameter $\alpha$, the transition rate from $m$ to $m-1$ is
\begin{equation}
\lambda_{a,\alpha}(m)
=
a r_0m
\exp\left(\frac{\alpha}{m}\right).
\end{equation}
Let $E_m$ be independent unit-rate exponential random variables. The waiting time in state $m$ can be represented as
\begin{equation}
\tau_m^{(a,\alpha)}
=
\frac{
E_m
}{
a r_0m\exp(\alpha/m)
}.
\end{equation}
Consequently,
\begin{equation}
\tau_m^{(a,\alpha)}
=
\frac{1}{a}
\tau_m^{(1,\alpha)}.
\end{equation}

If the cross-section begins with $k$ bonds, its failure time is $T_k^{(a,\alpha)}
=
\sum_{m=1}^{k}
\tau_m^{(a,\alpha)}$. Therefore,
\begin{equation}
T_k^{(a,\alpha)}
\overset{d}{=}
\frac{1}{a}
T_k^{(1,\alpha)}.
\end{equation}
In particular,
\begin{equation}
\mathbb{E}
\left[
T_k^{(a,\alpha)}
\right]
=
\frac{1}{a r_0}
\sum_{m=1}^{k}
\frac{
\exp(-\alpha/m)
}{m}.
\end{equation}

\subsection*{System-level scaling}

Let the state of the (intact) mesh be $\mathbf{k}
=
(k_1,\dots,k_{N_V})$. The total rate of the next rupture is
\begin{equation}
\Lambda_{a,\alpha}(\mathbf{k})
=
a r_0
\sum_{i}
k_i\exp\left(\frac{\alpha}{k_i}\right).
\end{equation}
Thus, $\Lambda_{a,\alpha}(\mathbf{k})
=
a\Lambda_{1,\alpha}(\mathbf{k})$. The probability that the next rupture occurs in cross-section $i$ is
\begin{equation}
\frac{
\lambda_{a,\alpha}(k_i)
}{
\Lambda_{a,\alpha}(\mathbf{k})
}
=
\frac{
k_i\exp(\alpha/k_i)
}{
\displaystyle
\sum_{j}
k_j\exp(\alpha/k_j)
}.
\end{equation}
This probability is independent of $a$. 
Since multiplying the hydrolase activity by $a$
multiplies every transition rate by the same factor,
the probabilities of the successive rupture events
remain unchanged. Now, each waiting time is multiplied by $1/a$ and, therefore the total time to system failure is also multiplied by $1/a$ giving

\begin{equation}
T_{\mathrm{sys}}^{(a,\alpha)}
\overset{d}{=}
\frac{1}{a}
T_{\mathrm{sys}}^{(1,\alpha)}.
\end{equation}
The corresponding survival function satisfies

\begin{equation}
S_{a,\alpha}(t)
=
S_{1,\alpha}(at).
\end{equation}
and also it is clear that
\begin{equation}
Q_q^{(a,\alpha)}
=
\frac{1}{a}
Q_q^{(1,\alpha)}.
\end{equation}

\subsection*{Long-cell limit}

In the long-cell limit, suppose that every cross-section begins with at least $m_*$ bonds and that a fraction $\pi_{m_*}$ begins with exactly $m_*$ bonds. At activity $a$, the early-time single-cross-section failure law is
\begin{equation}
\mathbb{P}\!\left(
T_{\mathrm{col}}^{(a,\alpha)}
\leq t
\right)
\sim
\pi_{m_*}
\exp(\alpha H_{m_*})
(a r_0t)^{m_*}.
\end{equation}
Therefore,
\begin{equation}
\mathbb{P}
\left(
T_{\mathrm{sys}}^{(a,\alpha)}>t
\right)
\approx
\exp\left[
-N_V\pi_{m_*}
\exp(\alpha H_{m_*})
(a r_0t)^{m_*}
\right].
\end{equation}

At reference activity and without load,
\begin{equation}
\mathbb{P}
\left(
T_{\mathrm{sys}}^{(1,0)}>t
\right)
\approx
\exp\left[
-N_V\pi_{m_*}
(r_0t)^{m_*}
\right].
\end{equation}

To make the loaded and unloaded survival probabilities
equal at the same time $t$, we equate their exponents.
This requires $a^{m_*}\exp(\alpha H_{m_*})=1$, giving the compensation factor
$a_{\mathrm{comp}}$ as
\begin{equation}
a_{\mathrm{comp}}
=
\exp\left(
-\frac{\alpha H_{m_*}}{m_*}
\right).
\end{equation}

\section{Numerical simulations}\label{app:simulations}

Stochastic simulations were performed in Python using NumPy, SciPy, and
Matplotlib. Unless otherwise stated, the initial number of peptide bonds in
each cross-section was sampled independently as
$K_i\sim\operatorname{Binomial}(L,p)$. Since the cross-sections evolve
independently as pure-death processes, their failure times were generated
separately. For a cross-section with $m$ intact bonds, the waiting time to the
next rupture was sampled from an exponential distribution with rate
\[
\lambda_\alpha(m)=r_0m\exp(\alpha/m),
\]
or $a\lambda_\alpha(m)$ when the relative hydrolase activity was $a$. These
waiting times were summed until the cross-section reached zero bonds, giving
its failure time $T_i$, and the system failure time was then obtained as
$T_{\mathrm{sys}}=\min_iT_i$. For $\alpha=0$, we used the equivalent
inverse-CDF sampler for the maximum of $K_i$ independent exponential bond
lifetimes. After rescaling time by $r_0$, these bond lifetimes have unit rate.
This procedure avoids explicitly simulating the sequence of rupture events
across the entire mesh.

As a numerical check, we compared this procedure with a direct event-driven
Gillespie simulation of the full mesh. For a current state
$\mathbf{k}=(k_1,\ldots,k_{N_V})$, the total rupture rate is
\[
\Lambda(\mathbf{k})=\sum_{i=1}^{N_V}\lambda_\alpha(k_i).
\]
The waiting time to the next rupture is sampled from
$\operatorname{Exp}(\Lambda)$, after which cross-section $i$ is selected with
probability $\lambda_\alpha(k_i)/\Lambda$. One bond is removed from the
selected cross-section, the rates are updated, and the procedure is repeated
until the first cross-section reaches zero bonds. Appendix Fig.~A.2 compares
the resulting system failure-time distribution with that obtained using the
independent-cross-section sampler for $N_V=5$, $L=40$, $p=0.4$, and
$\alpha=10$, using 3000 independent realizations for each method. The two
distributions closely overlap, with mean failure times $0.269$ and $0.273$,
respectively, and a two-sample Kolmogorov--Smirnov statistic $D=0.036$.
\clearpage

\begin{figure}[t]
\centering
    \includegraphics[width= 4 in]{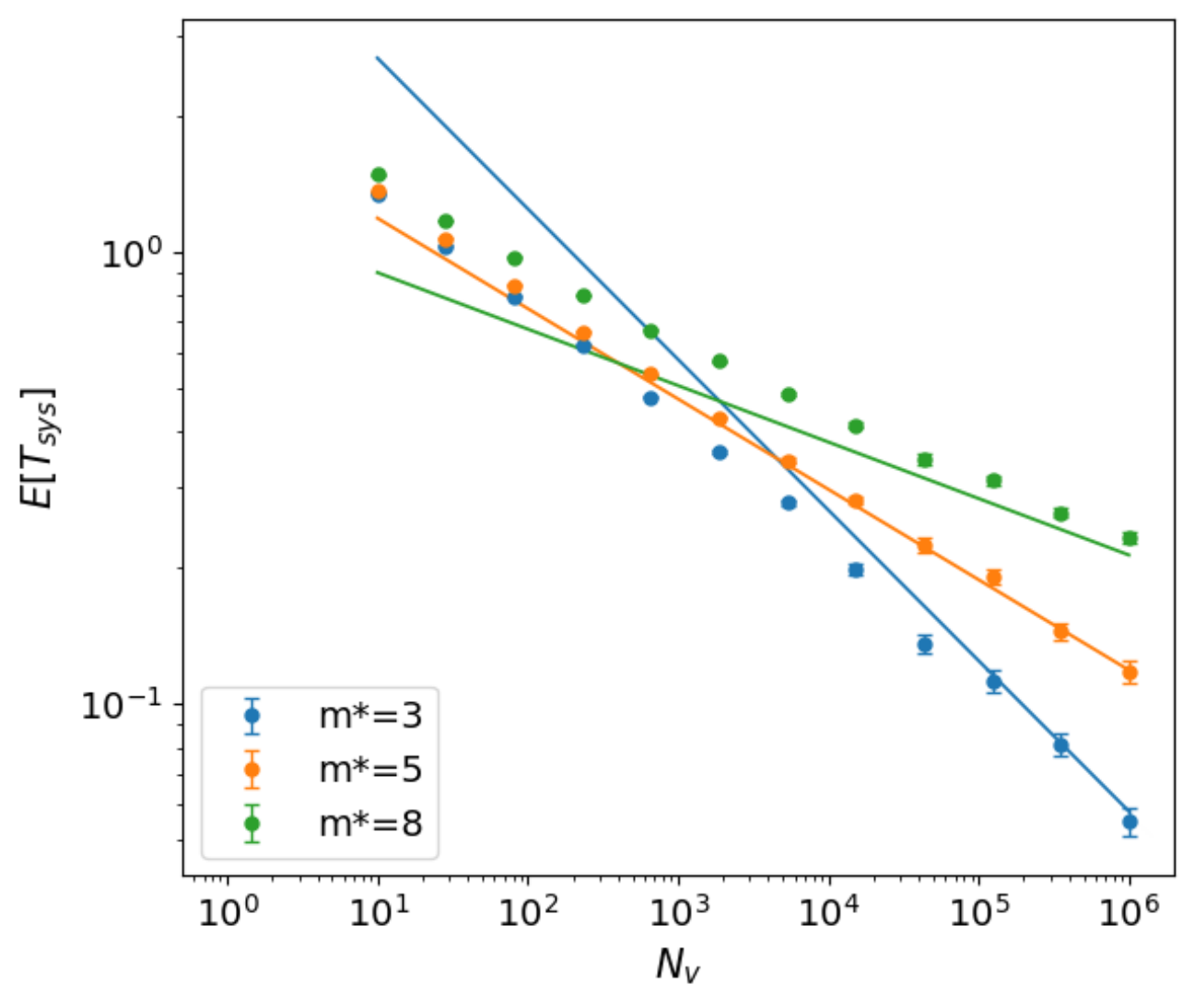}
     \caption{\textbf{Weakest-link scaling of system failure without mechanical loading.}
Mean system failure time, $E[T_{\mathrm{sys}}]$, as a function of the number of cross-sections $N_V$ for $L=40$, $p=0.25$, and minimum initial peptide bond numbers $m^*=3,5,$ and $8$. Initial bond numbers are sampled from a $\operatorname{Binomial}(L,p)$ distribution conditioned on $K_i\geq m^*$. Means are shown from 30--3000 independent system replicates per $N_V$ and $m^*$, with the number of replicates adjusted with $N_V$ to control the total number of simulated cross-sections; error bars denote $\pm$ SEM. Solid lines show the large-$N_V$ prediction $E[T_{\mathrm{sys}}]\propto N_V^{-1/m^*}$.}
    \label{fig:appendix_fig1}
\end{figure}

\begin{figure}[t]
\centering
    \includegraphics[width= 4 in]{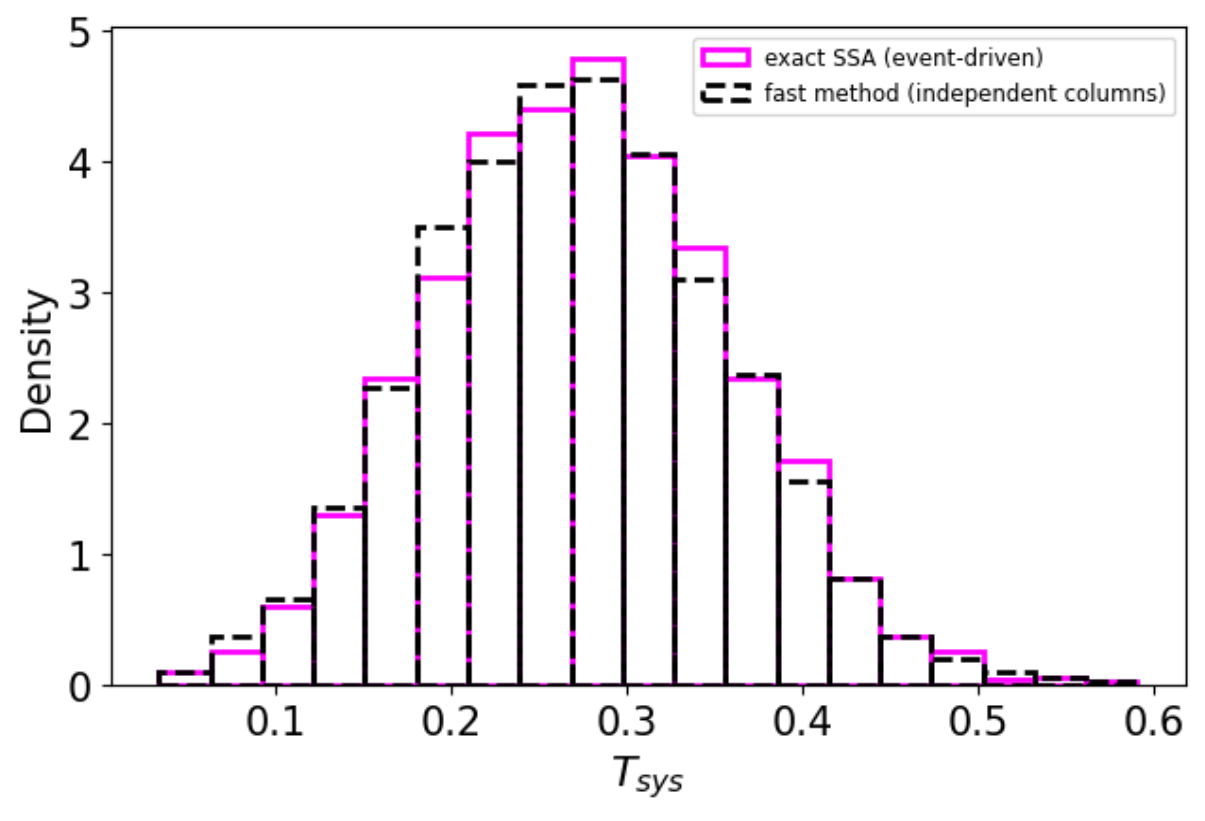}
     \caption{\textbf{Validation of the fast simulation method.}
Comparison of the vectorized independent-column method with a full event-driven Gillespie
simulation of the multi-column process for $N_V = 5, L = 40, p = 0.4$, and $\alpha = 10$. The two methods give closely overlapping failure-time distributions, with mean failure times $0.269$ and $0.273$ for the event-driven and vectorized simulations, respectively, and a two-sample Kolmogorov--Smirnov statistic $D=0.036$ ($n=3000$ realizations per method).
}
    \label{fig:appendix2}
\end{figure}

\begin{figure}
\centering
    \includegraphics[width= 5 in]{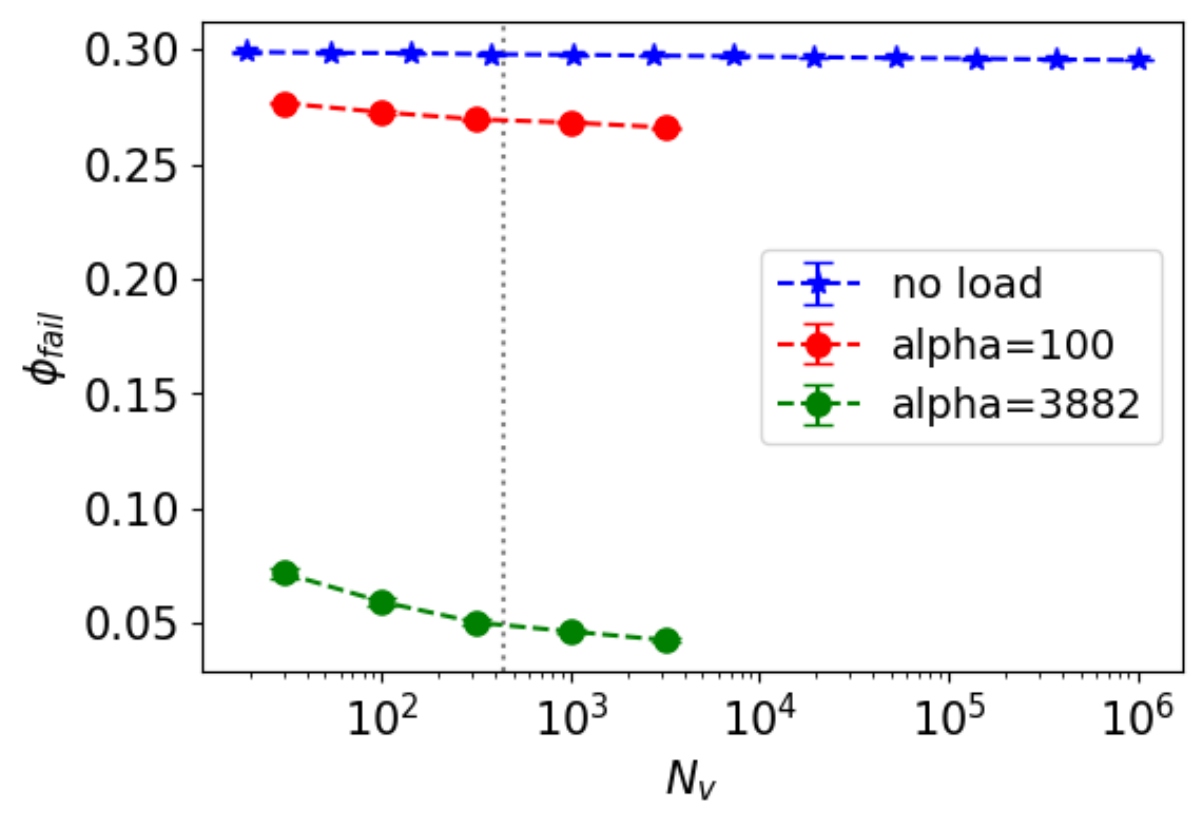}
   \caption{\textbf{Finite-size dependence of $\phi_{\mathrm{fail}}$ on axial system size $N_V$.}
Mean $\phi_{\mathrm{fail}}$ is shown as a function of the number of cross-sections $N_V$, with $L=3050$ and $p=0.30$ fixed, for no load ($\alpha=0$), moderate load ($\alpha=100$), and physiological load ($\alpha\simeq3882$). The vertical dotted line marks the physiological estimate $N_V=444$. Points show ensemble means over 60 realizations. $\phi_{\mathrm{fail}}$ depends only weakly on $N_V$ in the absence of load, whereas it decreases more appreciably with increasing $N_V$ under mechanical load, with the strongest finite-size dependence at physiological load.}
    \label{fig:appendix3}
\end{figure}
\label{app:failure_geometry}

\begin{figure}
\centering
    \includegraphics[width= \columnwidth]{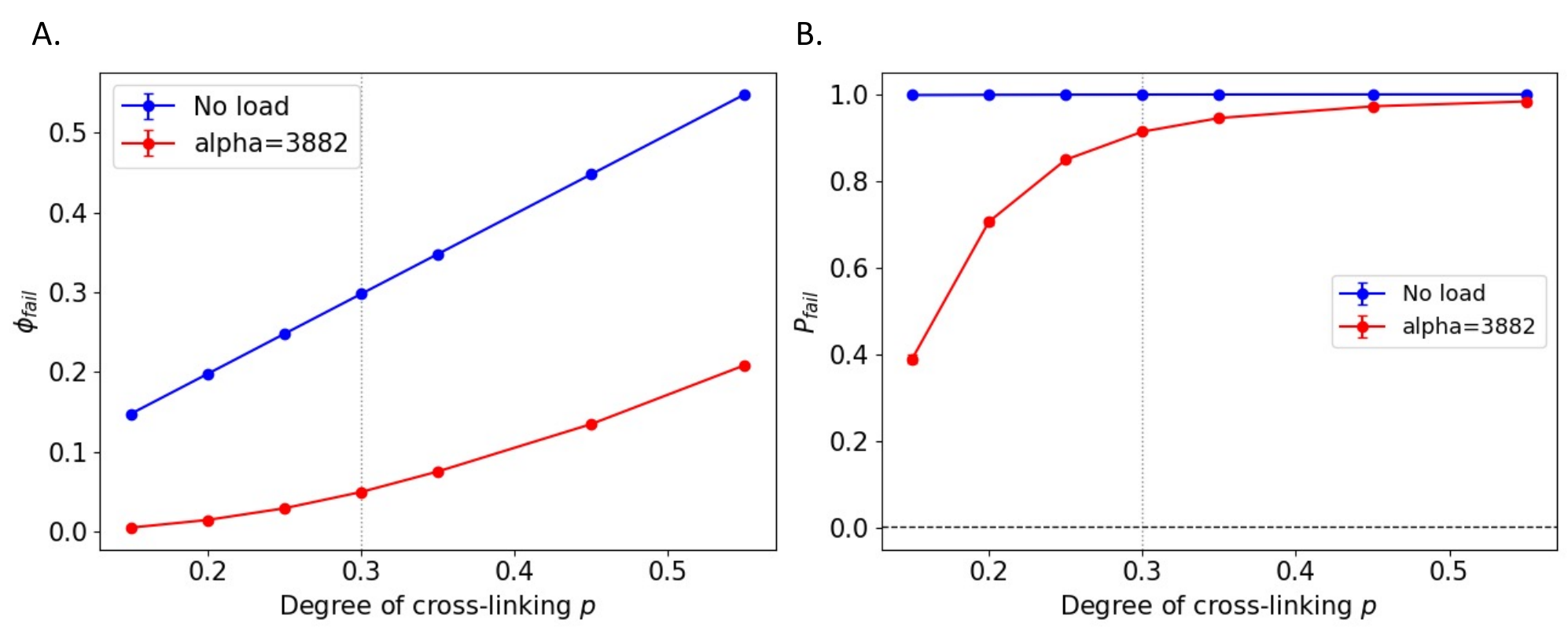}
    \caption{\textbf{Effect of cross-linking degree $p$ on failure geometry.}
Comparison of no load ($\alpha=0$) and physiological load ($\alpha\simeq3882$), with physiological parameters $L=3050$ and $N_V=444$ fixed; the vertical dotted line marks the physiological estimate $p=0.30$. Points show ensemble means over 150 realizations. \textbf{(A)} $\phi_{\mathrm{fail}}$ increases with $p$ in both regimes but remains substantially lower under load. \textbf{(B)} $P_{\mathrm{fail}}$ remains close to $1$ without load, whereas under load it increases from $\sim 0.39$ at $p=0.15$ to $\sim 0.98$ at $p=0.55$. The horizontal dashed line marks $P_{\mathrm{fail}}=1/N_V$, corresponding to the maximally localized limit. Thus, lower cross-linking produces more spatially localized damage at failure under load, a feature not captured by $\phi_{\mathrm{fail}}$ alone, which measures the total damage rather than its spatial distribution.}
    \label{fig:appendix4}
\end{figure}

\clearpage
\bibliographystyle{unsrt}
\bibliography{paper}

\end{document}